\documentclass[useAMS,usenatbib]{mnras}
\usepackage{graphicx}
\usepackage{amsmath} 
\usepackage{txfonts}
\usepackage[utf8]{inputenc}
\usepackage{aas_macros}
\usepackage{hyperref}

\usepackage{soul}  
\usepackage{xcolor}

\title[Determination of the turbulent parameter $\alpha$ in 4U\,1543$-$47 
(2002)]{
Determination of the turbulent parameter in {accretion} disks:
effects of self-irradiation in 4U\,1543$-$47 during the 2002 outburst}
\author[Lipunova, Malanchev]{
\parbox{\textwidth}{\raggedright G. V. 
Lipunova$^1$\thanks{E-mail:galja@sai.msu.ru}, K. L. Malanchev$^{1,2}$}
\\
\\
$^1$ Moscow  Lomonosov State University, Sternberg Astronomical Inst., 
Universitetski pr. 13, Moscow 119234, Russia\\
$^2$ Moscow Lomonosov State University, Faculty of Physics, Leninskie Gory, 1, 
Moscow 119234, Russia}

\begin{document}
\def\ak{a_\mathrm{Kerr}}
\def\at{\color{red}}
\def\acold{\alpha_\mathrm{cold}}
\def\ahot{\alpha_\mathrm{hot}}
\def\col{\mathrm{col}}
\def\fc{f_\col}
\def\di{\mathrm{d}}
\def\eff{\mathrm{eff}}
\def\grav{\mathrm{grav}}
\def\h{\mathrm{h}}
\def\irr{\mathrm{irr}}
\def\in{\mathrm{in}}
\def\mag{\mathrm{mag}}
\def\Msun{\mathrm{M}_{\odot}}
\def\Rsun{\mathrm{R}_{\odot}}
\def\nut{\nu_\mathrm{t}}
\def\Porb{P_\mathrm{orb}}
\def\rhot{R_\mathrm{hot}}
\def\Thot{T_\mathrm{hot}}
\def\Tin{T_\in}
\def\out{\mathrm{out}}
\def\vis{\mathrm{vis}}
\def\tvis{t_\vis}
\def\sigmaSB{\sigma_\mathrm{SB}}
\def\src{4U\,1543$-$47{}}
\def\x{\mathrm{x}}
\def\const{\mathrm{const}}
\def\grams{g\,s$^{-1}$}
\def\kms{km\,s$^{-1}$}


\pagerange{\pageref{firstpage}--\pageref{lastpage}} \pubyear{2000}

\maketitle

\label{firstpage}

\begin{abstract}
We investigate the viscous evolution of the accretion disk in \src, a black hole binary 
system, during the first 30 days after the peak of the 2002 burst by comparing {the} 
observed and theoretical accretion rate evolution $\dot M(t)$. The observed $\dot M(t)$ is 
obtained from spectral modelling of the archival RXTE/PCA data. Different scenarios of 
disk decay evolution are possible depending on {a} degree of  self-irradiation of the disk 
by the emission from its centre. If the self-irradiation, which is parametrized by factor 
$C_\irr$,  {had been} as high as $\sim 5\times10^{-3}$,  then the disk {would have been} 
completely ionized up to the tidal radius and the short time of the  decay {would have 
required} the turbulent parameter  $\alpha\sim 3$.  {We find that the shape of the $\dot 
M(t)$ curve is much better explained in a model with a shrinking high-viscosity zone. If 
$C_\irr\approx(2-3)\times 10^{-4}$, the resulting $\alpha$ lie in the interval $0.5-1.5$ 
for the black hole masses in the range $6-10~\Msun$, while the radius of the ionized disk 
is variable and controlled by irradiation.  For very weak irradiation, $C_\irr < 1.5 
\times10^{-4}$, the burst decline develops as in normal outbursts of dwarf novae with 
$\alpha \sim 0.08-0.32$.} The optical data indicate that $C_\irr$ in \src~(2002)  was not 
greater than {approximately} $(3-6)\times10^{-4}$. {Generally, modelling  of  an X-ray 
nova burst allows one to estimate $\alpha$} that depends on the black hole parameters.  
{We present} the public 1-D code {\sc freddi}  to model the viscous evolution of {an} 
accretion disk. Analytic approximations  are derived to estimate $\alpha$ in X-ray novae 
using $\dot M(t)$. 
\end{abstract}


\section{Introduction}

{The viscosity in the accretion disks is explained by the magnetohydrodynamic 
(MHD) turbulence, developing due to the magneto-rotational instability 
\citep{Hawley-Balbus1991,velikhov1959,chandra1961}. MHD simulations of the turbulence have 
been generally providing the turbulent parameter $\alpha \sim 0.01$. More recently,  numerical 
simulations  have achieved a value of $\alpha \sim 0.1$ for the regions with the partial ionization and convection \citep{hirose_et2014}.
Yet observations sometime indicate higher values: $\alpha \sim 0.1-0.4$~\citep{King_et2007},  
$0.5-0.6$~\citep{suleimanov_etal2008, malanchev-shakura2015}. 
 In this work,
  we study in detail an outburst of one X-ray nova 
  and derive constraints on $\alpha$.}

The thermal instability  in the zone of partial ionization of hydrogen in 
an accretion disk around a compact star is believed to be {a}  trigger mechanism 
of outbursts in  X-ray novae. The Disk Instability Model (DIM)   was originally 
put forward to explain  dwarf novae outbursts \citep[see reviews 
by][]{smak1984c,lasota2001}. The accumulation of mass in the cold disk 
(where the effective temperature is $\lesssim 10^4$~K and the matter is not 
ionized) between bursts leads to the situation when the surface density 
exceeds a critical value, and {the transition} to the hot state occurs on the thermal 
timescale. As the model predicts, this happens first at some radius, {proceeding then  to other disk rings outwards and inwards as an
`avalanche' and converting the 
disk to the hot state}. The viscous evolution of the disk leads to an {increase} of the central 
accretion rate, which is observed as an X-ray outburst if the central body is a 
compact object. To  reproduce  outburst cycles of dwarf and X-ray novae, {the $\alpha$-parameter  should be} higher in the hot state {by 1--2 orders of magnitude  than in the cold 
state}~\cite[see, 
e.g.,][]{lasota2001,King_et2007}.

\begin{figure*}
\rotatebox{0}{{\resizebox{0.98\textwidth}{!}
{\includegraphics{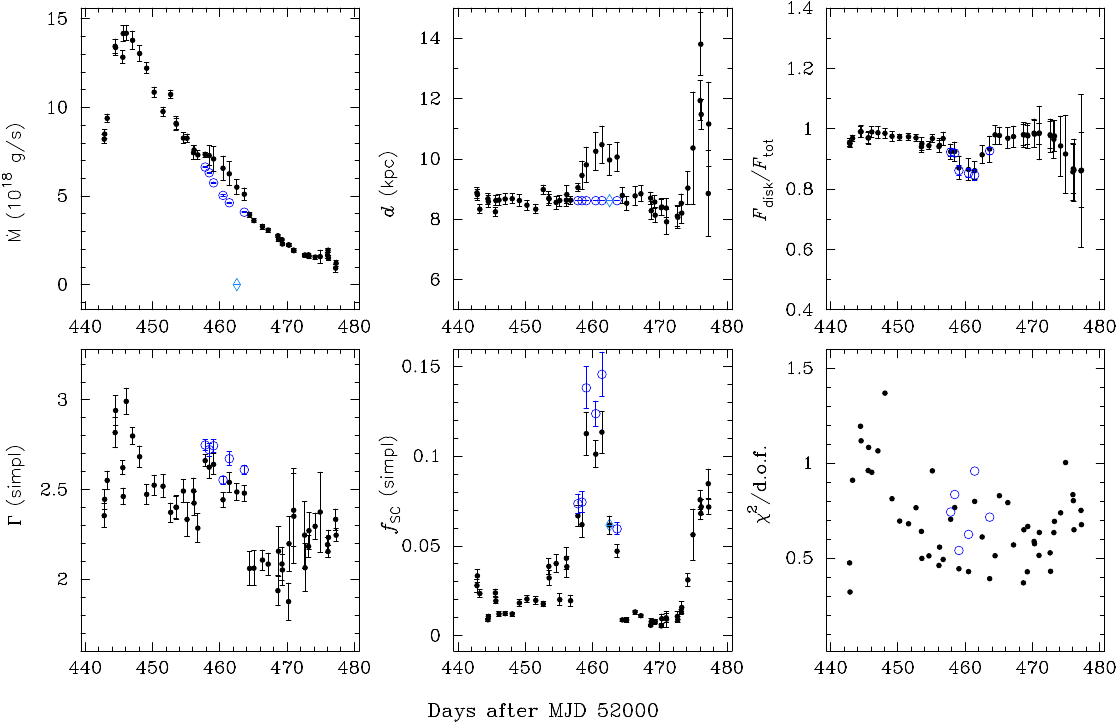}
}}} 
\caption{
Evolution of spectral  parameters $\dot M$, $d$, $\Gamma$ and 
$f_\mathrm{SC}$ in  {\sc XSPEC} fits {made with  the} model $tbabs * (simpl*kerrbb+laor)*smedge$  
{for the} 2002 outburst of \src~ observed by RXTE/PCA. {The} BH parameters are 
$m_\x=9.4$ and $\ak=0.4$, {the} disk inclination {is}
 $i=20\fdg7$.  {The left panels show the  disk-to-total-flux ratio} 
 in $0.5-50$~keV and reduced $\chi^2$. Empty  circles denote the spectral fits made with the fixed distance  
 $\approx 8.62$~kpc, which has been
 found as the average for {the observations at moments when the distance is least variable} 
 (before MJD 52456).  {The}  diamond stands for  {the} spectral fit with  {a value of the} reduced $\chi^2 
>2$.
  }
\label{fig.sp_fit_res}
\end{figure*}

 The central accretion rate during a burst is determined by the viscous 
evolution in the high-viscosity hot part of the disk because the cold part 
evolves at a relatively  {low} rate.  {The analysis} of the viscous evolution of X-ray 
novae  during outbursts {provides an} estimate of the turbulent parameter $\alpha$ 
in the hot state. The size of the hot part of the disk depends on the radial 
temperature distribution, meanwhile the disk temperature  depends on the viscous heating 
rate  {and} the incident flux of central X-rays. Previously, 
\citet{suleimanov_etal2008} have analysed X-ray and optical light-curves  of 
A\,0620-00 (1975) and GS\,1124-68 (1991), which have  fast-rise 
exponential-decay (FRED) X-ray light curves.  {They have considered the completely ionized disks with the
fixed outer radius and obtained values of $\alpha$ depending on the black hole 
parameters (see also \citet{malanchev-shakura2015})}.

In the present work, we study details of the viscous evolution of the disk in  
X-ray transient \src~during the first $\sim 30 $~days of  {the} decay of its FRED-like  
outburst in 2002.  {The} archival X-ray spectral observations by  RXTE/PCA are used 
to derive the evolution of the central accretion rate $\dot M(t)$. 
 System \src~(V*~IL~Lup) is a low mass X-ray binary  (LMXB) that 
shows outbursts about every ten years. The compact accreting object 
is a reliable black hole (BH) candidate. The binary has {an} orbital period {of}
$1.116$~day~\citep{orosz2003}, which is longer than those 
of most LMXBs with known orbital periods~\citep{rkcat}. For example, the   X-ray novae  with  
BH candidates  A~0620--00, GS~1124--68, GS~2000+25, {and } GRO~0422+32 have 
orbital periods less than 10 hours. The optical companion  {to} \src~is an A2V 
star with {a } mass {of } $M_\mathrm{opt} \sim 2.5~\Msun$. The mass of the black hole {has been}
estimated as $M_\x=2.7 - 7.5~\Msun$ or $M_\x = 9.4\pm 
2~\Msun$,  and the distance $9.1 \pm 1.1$~kpc~\citep{orosz_et1998, orosz_et2002}. The latest observational result is 
 a BH mass of $M_\x = 8.4- 10.4~\Msun$  and an inclination of $20\fdg 7\pm 1\fdg 
5$ for the binary system~\citep{orosz2003}.

 Different scenarios or stages of  {the} disk evolution during an outburst of an X-ray Nova  are possible, which {generally}
follow each other.  {During some time,}  the size of the hot zone can be constant, {being approximately equal to 
the tidal truncation radius in the binary system.} In case of \src, this 
requires a strongly irradiated accretion disk. {For less
irradiated disk, the hot zone should be shrinking, although  the irradiation may play} an 
important role in the disk evolution. Finally, { a stage of virtually negligible irradiation follows, resembling} the disk decay in dwarf novae.

{Using the observational data for \src, we study all these options.}
For a grid of the parameters $m_\x$ and $\ak$, we 
compare the accretion rate $\dot M(t)$ obtained from the RXTE data with the 
theoretical results. {The} optical light curves in $V$ and $J$  {bands} obtained by 
\citet{buxton_bailyn2004} are used to check the degree of self-irradiation. 
We assess the turbulent $\alpha$-parameter 
in the hot state, {as well as other disk properties of \src, and generalize our 
modelling results.}

To model the evolution of a hot irradiated disk, \hbox{1-D code} {\sc 
freddi} {has been developed and used in this} work. {The main 
input parameters of the code are as follows}: the turbulent parameter $\alpha$,  
{the} BH mass 
$m_\x$\footnote{We use two notations for the black hole mass: $M_{\x}$ is the 
mass in grams and $m_\x  = M_\x / \Msun$ is the mass in solar masses.}, {and} Kerr 
parameter $\ak$. For a non-irradiated disk, we expect that a cooling front, {which resembles that 
in a dwarf-nova disk,} controls $\dot M(t)$ and the X-ray light curve. To model such 
evolution, we follow the approach used by \citet{kotko-lasota2012} for normal 
outbursts of dwarf novae.


The structure of the paper is as follows. The spectral modelling of RXTE 
archival data {used to determine the} accretion rate evolution  is outlined in 
\S\ref{sec.sp_fit}. In \S\ref{sec.find_alpha} we describe the ways of 
determining $\alpha$ for irradiated and non-irradiated disks. In 
\S\ref{sec.results} the modelled light curves and $\alpha$ are presented. 
The optical emission during the outburst is considred in \S\ref{s.optical_emission}.
We discuss {the self-irradiation } of \src~and the model assumptions in 
\S\ref{sec.disc} and give a summary in \S\ref{sec.summary}. All statistical 
errors in the work are {the} 1-$\sigma$ confidence intervals.

\section{Spectral modelling  {of the} accretion rate evolution of \src~in 
2002}\label{sec.sp_fit} 

The outburst of \src~in 2002  {had} a FRED-like X-ray light curve with the 
first minor X-ray/IR re-flare at  $\sim 15$~day and the second bigger IR/optical 
re-flare  at  $\sim 40$~day {from the peak}~\citep{park_etal2004,buxton_bailyn2004}. We  analyse 
the same  data    obtained  with the Proportional Counter Array aboard the {\em 
RXTE} observatory as   {in} \citet{park_etal2004}. 

 There are two main spectral components in the \src\, burst spectra, as 
found  by \citet{park_etal2004}:   the multicolour blackbody-disk thermal 
emission  with a maximum around 1~keV and  {the} non-thermal component at higher 
energies. To describe the thermal emission of the disk,  {\citet{park_etal2004} used  the
$diskbb$ model} as a spectral component in {\sc XSPEC}~\citep{arnaud1996}. In this 
model,  parameter $ T_\mathrm{in}$, the `inner' disk temperature, is an 
indicator of the accretion rate through the inner edge of the disk.  {An} actual
value of the accretion rate is related to $\Tin$ in a complex way because of 
the general relativity effects.

We use {the} $kerrbb$ spectral model \citep{li_et2005} in {\sc XSPEC} that  takes 
into account general relativity effects on flux production and {
propagation of photons} in the vicinity of {a black hole, with $\dot M$ as a free 
parameter.} Here we outline the principle features of our modelling, and the 
details can be found in Appendix~\ref{a.sp_mod}.

\begin{figure}
\rotatebox{0}{{\resizebox{0.45\textwidth}{!}
{\includegraphics{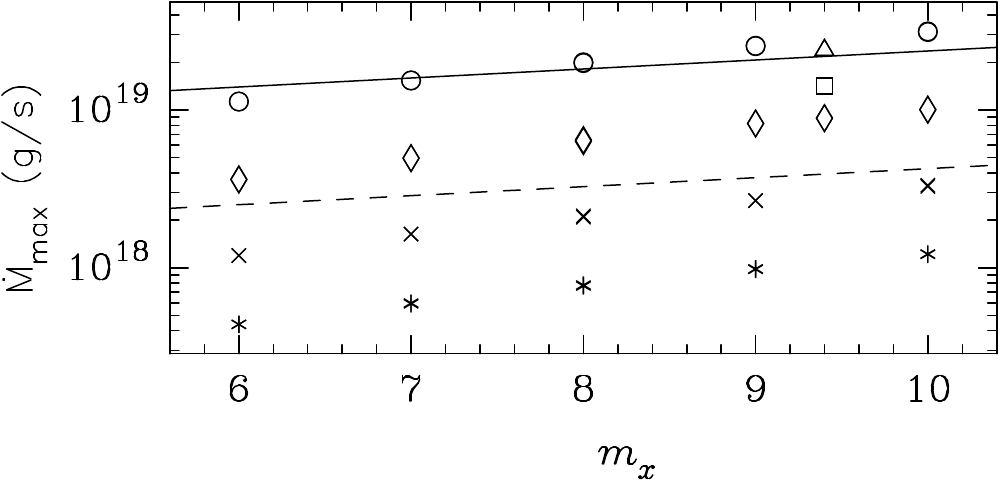}
}}} \hskip 0cm
\caption{
{The modelled} peak accretion rate of {the} 2002 outburst of \src~ (at $\sim$ MJD 52446) in 
{different} spectral models  with $i=20\fdg7$ and $f_\col=1.7$ (see the caption 
of Table~\ref{tab.parameters}) versus the BH mass. {The results} for both 
spectral combinations are plotted, {although the difference of their} peak accretion rates 
is {rather} imperceptible. The symbols indicate {the different} values of $\ak$: 0  
(circles), 0.1 (triangle), 0.4 (square), 0.6 (diamonds), 0.9 (crosses), and 
0.998 (asterisks). Two straight lines show  the  accretion {rates}  corresponding 
to the Eddington limit $L_\mathrm{Edd}$ and different { efficiencies of 
the accretion} onto the black hole: for $\ak=0.0$ (solid) and $\ak=0.998$ 
(dashed). See Appendix~\ref{a.sp_mod} for details.
   }
\label{fig.dotMpeak}
\end{figure}

  The $kerrbb$  parameters, which we set free, are the accretion rate $\dot M$ 
and {the} distance $d$. {For each fit,} we fix the BH mass $M_\x$ and {the} dimensionless Kerr 
parameter $\ak$. The disk inclination is the binary orbit inclination,  $20\fdg7$~\citep{orosz2003}, or, 
alternatively, $32\degr$~\citep[{a value suggested by spectral modelling;}][]{morningstar-miller2014}.  The list of 
these and other parameters can be found in Table~\ref{tab.parameters} in 
Appendix~\ref{a.sp_mod}.

  To describe the non-thermal component as the thermal emission comptonized by 
the high-temperature plasma near the disk, we  use  the convolution model  
$simpl$~\citep{steiner_et2009} or the additive model 
$comptt$~\citep{titarchuk1994}. Thus, two combinations of spectral components 
are used, including either $simpl$ or $comptt$.   As we {have found}, both 
spectral combinations give very similar peak magnitudes of $\dot M$.
  
For each spectral combination, we {have obtained} at least 20 accretion rate curves: for 
each pair of values of the BH mass and {the} Kerr parameter. An example of {the}
evolution of some spectral parameters is shown  in Fig.~\ref{fig.sp_fit_res}, 
where we adopt observationally suggested (hereafter `central') disk 
parameters: $m_\x=9.4$, $a=0.4$, $i=20\fdg7$~\citep{orosz2003}.

The peak accretion {rates} versus the BH parameters are shown in 
Fig.~\ref{fig.dotMpeak}. The higher $\ak$,  the less the accretion rate is
needed to generate {the} observed X-ray flux. The obtained 
accretion {peak rates} {for $\ak=0$}  are very close to the critical Eddington rate 
calculated as $L_\mathrm{Edd}/(\eta(\ak)\,c^2) = (4\,\piup\, G\, M_\x\, 
m_\mathrm{p})/(c\, \sigma_\mathrm{T}\, \eta(\ak))$, where $\eta(\ak)$ is the 
accretion efficiency. Above the limit, the standard thin disk model is 
inaccurate. However, we neglect this possibility in the present study, 
regarding the limit as a formal approximate value, meanwhile the excess is quite 
small.

For {the} observations around  MJD 52460, when the accretion rate curve {shows} a 
bump (see the top left panel of Fig.~\ref{fig.sp_fit_res}), we have made 
additional spectral fits, fixing the distance at its average value (it is found 
as the average from the spectral fit results before MJD 52456 if $\chi^2<2$). 
 The additional fits demonstrate that, apparently, the accretion rate  decays 
 smoothly (empty circles in Fig.~\ref{fig.sp_fit_res}). The 
bump on the $\dot M(t)$ curve (solid circles around MJD 52460) is probably a 
result of strong variations of the non-thermal component, or {a manifestation of an} extra spectral component, {which cannot be properly described by the presumed 
spectral model.}

\section{Determination of the value of $\alpha$ from X-ray nova outbursts} 
\label{sec.find_alpha}
{The} viscous evolution of an accretion disk {involves} the radial redistribution of the 
viscous torque and {the} mass density. This process is described by the equation of diffusion type: 
\begin{equation}
    \frac{ \partial \Sigma }{\partial t} = \frac{1}{4\piup} \frac{(G 
M_\x)^2}{h^3} 
\frac{\partial^2 F}{\partial h^2},
\label{eq.diffusion}
\end{equation}
where $h \equiv \sqrt{G M_\x r}$ is the specific angular momentum, $\Sigma$ is 
the surface density, $F=2\,\piup\,W_{r\varphi}\,r^2$ is the viscous torque 
expressed {using} the height-integrated viscous stress tensor 
$W_{r\varphi}$~\citep[see, e.g.][]{lyub-shak1987}. The viscous torque and 
surface density in the Keplerian disk are related as follows: 
\begin{equation}
F = 3\,\piup\, h\, \nut\, \Sigma\, ,
\label{eq.Sigma_F} 
\end{equation}
where $\nut$ is the kinematic coefficient of turbulent viscosity.

Using $F(h)$ as a radial characteristic  is very effective. First, boundary
conditions  are usually set  on $F$ or its first derivative $\partial F / 
\partial h  = \dot M$.
Second, the viscous heating in the disk is explicitly related to the viscous 
torque, $Q_\vis = 3\, (G\,M_\x)^4\,F/(8\,\piup\,h^7$), and
this relation holds for cases with $\dot M=0$ too.

\begin{figure}
 \vskip -1cm
 \rotatebox{0}{{\resizebox{0.44\textwidth}{!}
 {\includegraphics{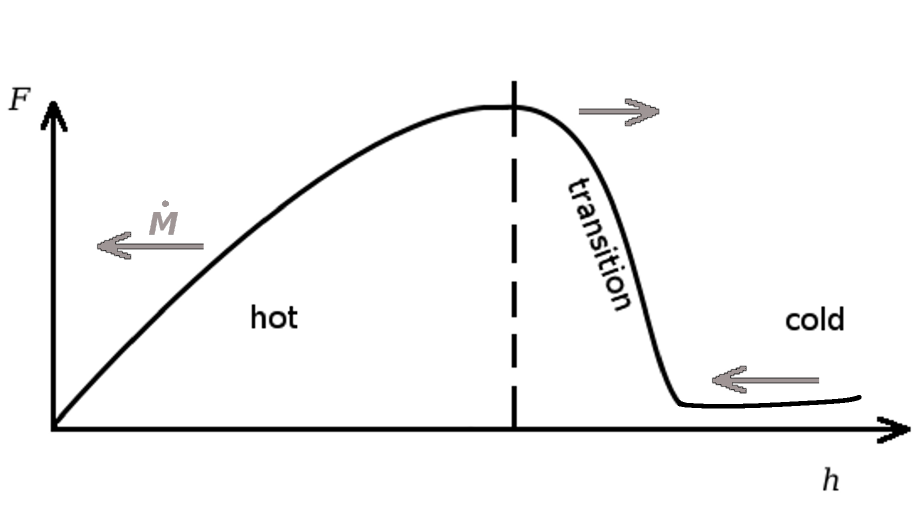}}}} \hskip 0cm
 \caption{
 A schematic representation of the viscous torque distribution in the disk, 
which has the hot inner and cold outer zone.  The grey arrows show {the direction of the mass flow}, and their heights indicate the
 representative moduli of the accretion rate. 
{The} slope of the curve $F(h)$  equals the accretion rate.  In the outer cold disk, the accretion rate 
is suppressed due to {a lower} value of the kinematic coefficient of turbulent viscosity.
 There are two 
locations where the accretion rate $\dot M = \partial F /\partial h$ {becomes} zero. 
} 
 \label{fig.scheme}
 \end{figure}

 To illustrate how the viscous torque behaves, we plot schematically {the}
$F(h)$-distribution in the disk with the hot 
inner and cold outer  parts (Fig.~\ref{fig.scheme}). 
At the inner  boundary of the disk  around {the} black hole,
the viscous torque is zero: $F_\in = 0$.
A positive slope corresponds to mass inflow and vice versa.
In the outer, cold neutral disk, the viscous torque is depressed.
Between the cold and hot zone, {where the hydrogen undergoes recombination/ionization,} 
the opacity coefficient {varies strongly}. If the disk 
in a binary system is fully ionized during some {perod of} time, the {outer radius of the disk may be considered stationary, corresponding to the dashed line} ~\citep[{the corresponding} 
quasi-stationary distribution of $F$ is derived by ][]{lipunova_shakura2000}. {The} mass is lost from 
the disk through {its} inner boundary. When the accretion rate drops {to such level} that the 
surface density in the hot disk becomes less than a critical value, the {outside-in}
transition to the cold state begins~\citep{smak1984c}.

At the outer boundary of the hot zone (the dashed line in 
Fig.~\ref{fig.scheme}), the  viscous torque $F$ reaches {an} extremum. 
Subsequently, we assume that the {mass flow becomes zero at the boundary:} 
$\partial F / \partial h  = 0$. There is a mass outflow in the transition 
zone, {and} the mass accumulates in 
the cold zone while {the} transition region moves towards the centre.  In the cold 
zone, it takes  {a longer time for a quasi-stationary distribution} to develop, hence, 
the slope of $F(h)$ is not constant there.

 {There are three successive  stages of an outburst decay in X-ray novae as}
suggested by \citet{dubus_et2001}: 
(1)  the hot zone has {the} constant radius $\rhot$ {due to the} strong 
irradiation (this stage can be absent if the disk is too large);
(2) the cooling front begins to move,  {its position being determined by the} irradiation;
(3) the cooling front propagates with  {the} speed $\sim \ahot \, u_\mathrm{sound}$,{when the} 
irradiation is not important. Two first stages are possible only  if the 
irradiation is strong enough.  Fig.~\ref{fig.scheme} qualitatively 
describes each of these three stages.  In the three following 
subsections, we describe  methods to model these stages.

\subsection{{The} hot disk with {a} fixed outer radius}\label{s.rconst}

If the outer radius  is constant, the $e$-folding time of the  accretion rate 
viscous decay { in the disk around a stellar-mass black hole in a binary system is}
\begin{equation}
 t_\mathrm{exp} \approx 0.45 \, R_\mathrm{out}^2/\nut(R_\mathrm{out})\, 
\label{eq.texp}
 \end{equation}
The above 
relation is obtained as an approximation of the exact solution {for evolution of the disk whose kinematic viscosity} 
 $\nut(r)$ {does not change with time}~\citep{lipunova2015}. {Such a viscous disk with the fixed outer radius yields a  precisely exponential decay of} $\dot M(t)$.
 
 {In} $\alpha$-disks~\citep{shakura-sunyaev1973}, the kinematic 
coefficient of turbulent viscosity depends on the surface density and, {respectively,} on 
time: $\nut = \nut (r,t)$. The solution for the decay of {a} hot $\alpha$-disk 
with {a} fixed outer radius {has been found} by \citet{lipunova_shakura2000}.  The 
accretion rate decays as $\dot M \propto t^{-10/3}$ in  the regime of the 
Kramers opacity. Such {a} steep decay can  be observed in X-rays as an exponential 
decay. 

The analytic solution {may} be  applied {for} a time interval when  the 
whole disk is in the hot state. This can happen due to {a} high  viscous heating or {a} strong irradiating flux,  
keeping the whole disk in the completely ionized state.~\citep[][also 
\citet{meyer-meyer1984,meyer-meyer1990}]{king-ritter1998,Shahbaz_etal1998, 
esin_et2000,dubus_et2001}. \citet{ertan-alpar2002} {have suggested} that {the} radius $\rhot$  {may be constant, although  smaller
than the disk size,} if a specific height profile is developed with a stationary {boundary} between the hot and cold zone, 
beyond which the disk is shadowed from the central X-rays.

One can estimate $\alpha$ using~\eqref{eq.texp}. Let us parametrize the local 
viscous stress tensor in the equatorial plane of the disk as $w_{r\varphi} = 
\alpha \, P_\mathrm{c}$, where $P_\mathrm{c}$ is the total pressure 
\citep{shakura-sunyaev1973}. For the outer parts of the hot disk, the gas 
pressure dominates. On the other hand,  $w_{r\varphi} = \frac 32\, 
\omega_\mathrm{K}\,\nut\,\rho_\mathrm{c}$ (c.f. Eq.~\eqref{eq.Sigma_F}).  
Substituting $P_\mathrm{c}/\rho_\mathrm{c} = \Re\, T_\mathrm{c}/\mu$ {yields }
a relation between $\nut$ and $\alpha$: $\nut = \frac 23\, (\alpha/ 
\omega_\mathrm{K})\, (\Re\, T_\mathrm{c}/\mu)$. Following the work of 
\citet{ketsaris_shakura1998}, we use the relation between $T_\mathrm{c}$ and 
the half-thickness $z_0$, which is the depth where the optical thickness 
$\tau=2/3$ and $T=T_\mathrm{eff}$: $\Re\, T_\mathrm{c}/\mu = \omega_\mathrm{K}^2 
\, z_0^2/\Pi_1$; $\Pi_1$ is {a} dimensionless parameter depending on the total 
optical thickness of the disk in the vertical direction; {it has been  } calculated by 
\citet{ketsaris_shakura1998} and \citet{malanchev_et2017}. We arrive at the  
relation $\nut(r) = \frac 23\, (\alpha/ \omega_\mathrm{K})\, 
(\omega_\mathrm{K}^2 \, z_0^2/\Pi_1)$  and, { hence,}
\begin{equation}
 \alpha \sim 0.15
  \,\left(\frac{R_\mathrm{out}}{2\,\Rsun}\right)^{3/2}
  \left(\frac{z_0/R_\out}{0.05}\right)^{-2}  \left(\frac{M_\x}{10\,
  \Msun}\right)^{-1/2} 
  \left(\frac{t_\mathrm{exp}}{30^\di}\right)^{-1} \times \Pi_1\, ,
\label{eq.alpha}
\end{equation}
where $t_\mathrm{exp}$ is the $e$-folding of the accretion rate decay,
$z_0$ is the disk half-thickness near the outer radius, {and} $\Pi_1=5.5- 6$.
To estimate $\alpha$, one should substitute  {$z_0$ corresponding to} the peak of an X-ray 
nova outburst.

The main uncertainty in the above formula is the radius of the disk.
In addition, the evolution of the  half-height of the $\alpha$-disk leads to a 
 variation of the numerical factor in \eqref{eq.alpha}. 
{However, the  modelling} of the disk evolution can provide a self-consistent value of $\alpha$. This
was done for outbursts of A\,\hbox{0620--00} (1975) and GS\,\hbox{1124--68} (1991)
by \citet{suleimanov_etal2008}.  
Using our code {\sc freddi}\footnote{{\sc freddi} can be 
freely downloaded from the authors' web page 
\url{http://xray.sai.msu.ru/~malanchev/freddi/}.},
we find that for any fully-ionized viscously-evolving accretion  disk
\begin{equation}
 \alpha \approx 0.21\, \left(\frac{\rhot}{\Rsun}\right)^{25/16} \,
   \left(\frac{t_\mathrm{exp}}{30^\di}\right)^{-5/4}\,
   \left(\frac{\dot M_\mathrm{max}}{10^{18}\,\mathrm{g\,s}^{-1}}\right)^{-3/8}\,
   m_\x^{5/16}  \, ,
   \label{eq.f_kramers}
\end{equation}
for the Kramers opacity, or
\begin{equation}
   \alpha \approx 0.20 \, \left(\frac{\rhot}{\Rsun}\right)^{12/7} \,
   \left(\frac{t_\mathrm{exp}}{30^\di}\right)^{-9/7}\,
   \left(\frac{\dot M_\mathrm{max}}{10^{18}\,\mathrm{g\,s}^{-1}}\right)^{-3/7}\,
   m_\x^{2/7}\, 
   \label{eq.f_opal}
\end{equation}
for the OPAL approximation.   Power indexes in the above expressions 
are obtained {when} substituting the thickness of the disk  in \eqref{eq.alpha} by 
its analytic expression from~\citet{suleimanov_et2007e}. {An} accuracy of {the}  numerical factors is 5\%.
Details of the {{\sc freddi} code} can be found in Appendix~\ref{a.code}.

%

 Let us estimate  $\alpha$ for \src, {supposing that} the fast viscous evolution {has swept} the whole disk.
 The Roche lobe effective radius of the primary in \src~is 
$R_\mathrm{RL}\approx4.2-5.3 \Rsun$ \citep{eggleton1983}  for the BH 
 mass lying in the interval  $M_\x = 6-10~\Msun$. {The maximum} disk  radius 
in \src~ can be as large as  $\sim 4 \Rsun$, if the tidal interactions 
with the companion truncate the disk at $R_\mathrm{tid} \approx 0.8\, 
R_\mathrm{RL}$ \citep{paczynski1977, suleimanov_et2007e}.
{
{When substituting} the radius and other parameters  for \src~in \eqref{eq.f_kramers} or \eqref{eq.f_opal}, {namely,} $t_\mathrm{exp}=15^\di$, $m_\x=9.4$, $\dot M_\mathrm{max}=1.4\times10^{19}$~g/s~(from Fig.~\ref{fig.sp_fit_res}, the upper left panel),
 we obtain $\alpha \sim 3.2-3.3$, an unplausible value.  This 
result indicates that {the disk evolved in a different manner}.  
Most probably, during the burst of 2002  the outer parts of the disk in 
\src~remained cold. }

\subsection{Disks with $\rhot$ controlled by irradiation}\label{s.irradiation_dominated_evolution}

After the recombination starts, the radius of the hot disk decreases. If 
the X-ray heating is strong, the radius of the hot zone $\rhot$ is {determined} by 
the incident flux, that is, by the central accretion rate. 
\citet{dubus_et2001} argue that {the} transition between the hot and cold portion{s} of 
the disk in X-ray transients is determined by the position where 
$T_\irr=10^4$~K. This {is explained by the fact that} the cold {state exists only} for lower 
$T_\mathrm{eff}$~\citep{meyer-meyer1984,tuchman_et1990}.   {Some evidence} for 
such behaviour {was} found by \citet{hynes_etal2002} in the $1999-2000$
outburst of XTE\,J1859+226, {although} they suggest{ed} that the condition at the hot 
zone radius $\rhot$ {may be more complex}.

We adopt the following parametrization for the flux irradiating the outer 
disk~\citep{lyutyi-sunyaev1976,Cunningham1976}: 
\begin{equation}
    Q_\irr \,\equiv\, \sigmaSB\,T_\irr^4 =C_\irr\, \frac{L_\mathrm{bol}}{4 
\,\piup\, r^2},
\label{eq.C_irr_def}    
\end{equation}    
where $L_\mathrm{bol} = \eta\,(\ak)\,\dot M\, c^2$ is the bolometric 
flux,  $\eta\,(\ak)$ is the accretion efficiency, $\sigmaSB$ is the 
Stephan--Boltzmann constant,  and $C_\irr$ is the irradiation parameter.
We assume in the present study that $C_\irr=\const$ (see discussion
 in \S\ref{ss.disc_cirr}).

We calculate the radius of the hot zone $\rhot(t)$  from 
\begin{equation}
\sigma\, \Thot^4 = C_\irr \, \frac{\eta(\ak)\, \dot M_\in(t)\, c^2 }{4\, 
\piup\,\rhot(t)^2 }\, ,
\label{eq.rhot}
\end{equation}
where $\Thot =10^4$~K and $\dot M_\in(t)$ is the accretion rate through the 
inner edge of the disk.
\begin{figure}
\rotatebox{0}{{\resizebox{0.45\textwidth}{!}
{\includegraphics{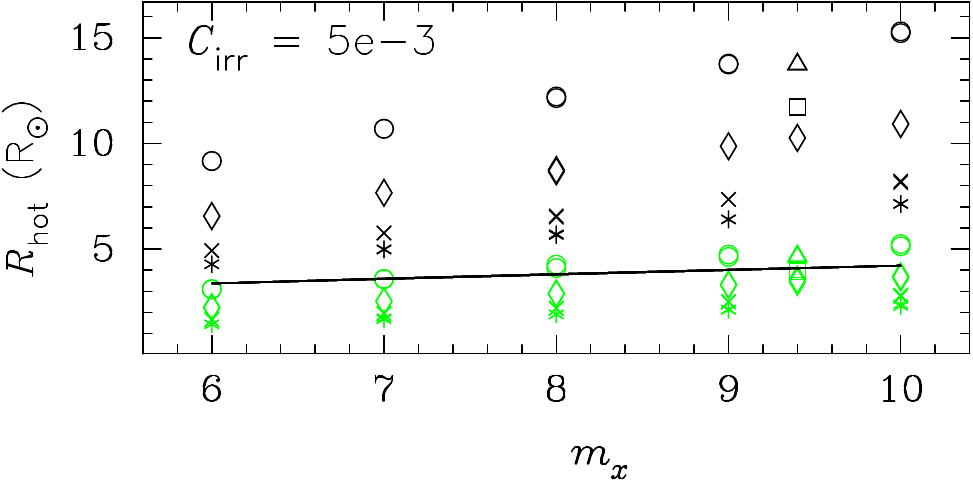}
}
}} \hskip 0cm
\rotatebox{0}{{\resizebox{0.45\textwidth}{!}
{\includegraphics{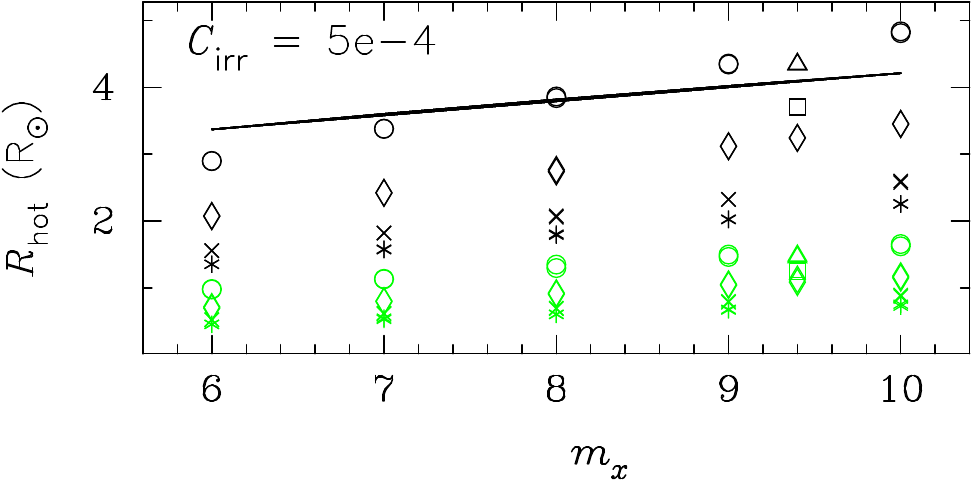}
}
}} \hskip 0cm
\caption{{The hot-zone radius ($T_\irr = 10^4$~K) as a function of BH mass  at the  accretion peak on} MJD 25446 (black
symbols), and at the end of the investigated light curve,  
MJD 25474 (green symbols online).  $C_\irr=5 \times 10^{-3}$ in the top panel,
and $C_\irr=5 \times 10^{-4}$ in the lower panel.
Symbols have the same meaning as in Fig.~\ref{fig.dotMpeak}, which shows the 
corresponding peak accretion rates.
Solid line is the tidal radius
of the accretion disk, which is a function of the central mass.  
}
\label{fig.Rpeak1}
\end{figure}

In Fig.~\ref{fig.Rpeak1}, we show {the hot zone size versus  the BH 
mass for \src~}, calculated for  two different irradiation factors: $C_\irr=5\times 
10^{-3} $ (the value suggested for X-ray novae \citep{dubus_et2001}) and  
$C_\irr=5\times 10^{-4}$. {The black symbols for} $\rhot$ correspond to  
the peak accretion rate {(MJD 25446)} and {the green symbols online represent the accretion rate} at the end 
of the studied interval  {(MJD 52474}).

When the accretion rate  is high and $\rhot(\dot M)>R_\mathrm{tid}$, the 
hot zone extends through {the} entire disk and, {hence},  the viscously evolving 
region has a constant radius. As we {can} see from Fig.~\ref{fig.Rpeak1} for $C_\irr=5\times 10^{-3} $ (top), 
the whole disk would have been hot during almost entire investigated time interval. For lower $C_\irr$,  the 
hot zone radius becomes {smaller} than $R_\mathrm{tid}$ (see Fig.~\ref{fig.Rpeak1}, 
the lower panel) and cannot be constant during the investigated time interval.

The relative importance of the irradiation  comparing to  the viscous heating can 
be expressed as follows:
\begin{equation}
\frac{Q_\irr}{Q_\vis} = \frac 43\, \eta(\ak)\, C_\irr\, \frac{r}{R_\grav}\, 
\frac{1}{f_F(r)}
\label{eq.irr_vis}
\end{equation}
\citep[e.g.][]{suleimanov_et2007e}, where $R_\grav = 2\,G\, M_\x/c^2$ and 
the viscous heating 
\begin{equation}
Q_\vis  =  \frac38 \frac{\sqrt{GM_\x}}{\piup} \frac{F}{r^{7/2}}\, , \qquad F = 
\dot M_\in\, h \, f_F(r) \,.
\label{eq.qvis}
\end{equation}
For the quasi-stationary solution of Eq.~\eqref{eq.diffusion}, {the dimensionless factor} 
$f_F(r) \approx 0.7$ at the radius where $\dot 
M=0$~\citep{lipunova_shakura2000}. The ratio \eqref{eq.irr_vis} does not depend 
on the accretion rate in the approximation of constant $C_\irr$. The 
illumination is more important for bigger disks. If $Q_\irr > Q_\vis$ at 
$\rhot$,
{a change  in} $\rhot(t)$ is {determined} by {a}  variation of the irradiating flux. As 
the latter depends on $\dot M(t)$, the 
radius of the hot zone shifts at a rate {determined} by the viscous timescale at 
$\rhot$.

\begin{figure}
\rotatebox{0}{{\resizebox{0.45\textwidth}{!}
{\includegraphics{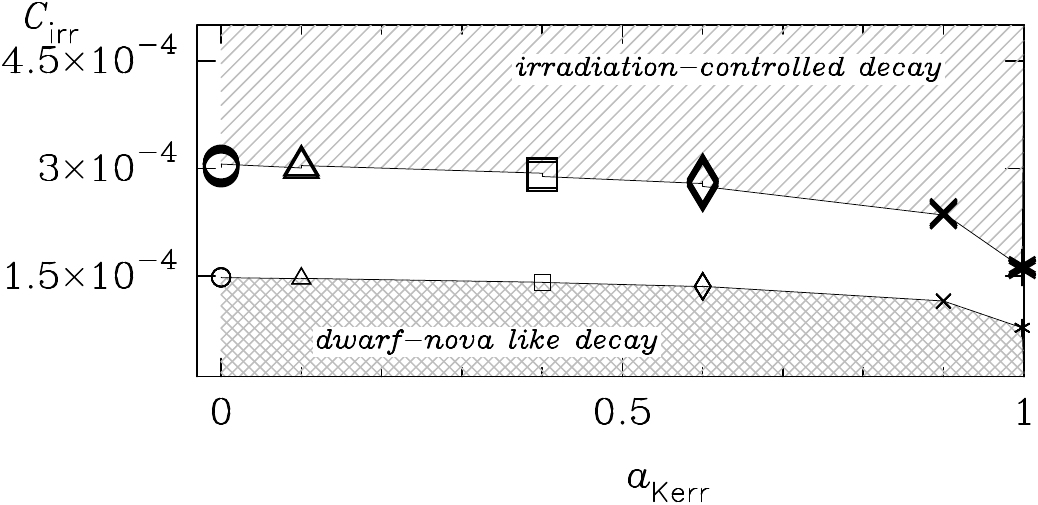}
}}} \hskip 0cm
\caption{Critical values of  $C_\irr$, determining the type of hot disk 
evolution in \src~during the 2002 burst, versus the Kerr parameter. The {points of the lower 
sequence} (small symbols) {represent} the boundary below which the 
irradiation is so low that it cannot influence the hot zone size, and the 
cooling front propagates like in a normal outburst of a dwarf nova.  Such $C_\irr$
are obtained from $Q_\irr/Q_\vis (\rhot) =1  $ at the burst peak.  The {big symbols in the higher  
sequence represent} the values of $C_\irr$,  above which the whole 
modelled light curve {should} be explained by evolution of the irradiated hot 
zone, {{following the equation} $Q_\irr/Q_\vis (\rhot) =1  $ at the end of 
the studied time interval}. { The symbols are of the same meaning} as in Fig.~\ref{fig.dotMpeak}. There is no 
dependence on $m_\x$. 
}
\label{fig.Cirr_ak}
\end{figure}

  Fig.~\ref{fig.Cirr_ak} presents values of $C_\irr$, critical in the 
context of {explaining} the decay of 2002 burst of \src. The  higher sequence  of symbols
represents the minimum values of $C_\irr$  above which the whole modelled 
light curve {should} be explained by the irradiation-controlled  hot zone. These 
values are found from {the}  condition $Q_\irr/Q_\vis =1  $ at  $\rhot$ at the end of 
the studied time interval, taking into account Eq.~\eqref{eq.rhot}.

To model evolution of the irradiation-controlled hot disk, we solve the 
viscous diffusion equation~\eqref{eq.diffusion}  on the interval from $R_\in$ 
to $\rhot$  using our {{\sc freddi} code}. {The condition $\dot M = 0$ is set at $\rhot$}. The outer boundary thus corresponds to   the vertical 
dashed line in Fig.~\ref{fig.scheme}. Physically, the transition to the cold 
disk starts there, so we  {determine} $\rhot$ at each step {using} \eqref{eq.rhot}.

 All models, calculated by {\sc freddi} for the present study, are obtained 
with the quasi-stationary distribution as  the initial one. This 
corresponds to  {the} calculation of the decaying part of the burst, although {\sc 
freddi} can simulate the light curve from the early rise~(see Appendix~\ref{a.code}).

\subsection{Cooling front without irradiation}\label{s.cf}

 If X-ray heating is very low, a so-called cooling front, surrounding   the 
hot zone,  propagates towards the centre while the accretion rate decreases. 
{The} heating and cooling fronts without 
irradiation have been  {studied } to explain dwarf nova 
bursts~\citep{hoshi1979, meyer_meyer-hofmeister1981, smak1984a, 
meyer1984,Lin_et1985, Cannizzo1994, 
vishniac-wheeler1996,menou_etal1999fronts,smak2000}.

Numerical modelling of DIM in dwarf novae shows that, if $\ahot$ is constant, 
the speed of the cooling front {approaches}  a characteristic constant velocity of 
order of $ \ahot\, 
u_\mathrm{sound}$~\citep{meyer1984,Cannizzo1994,vishniac-wheeler1996}, {with the 
hot inner part of the disk evolving} in a self-similar 
way~\citep{menou_etal1999fronts}.

Using analytic approximations to numerical results of DIM, 
\citet{kotko-lasota2012}  {considered} about 20 outbursts of dwarf novae and AM CVn 
stars  to determine the hot disk viscosity parameter $\ahot$ from the measured 
decay times. This method relies on the implied velocity of the cooling front 
propagation. \citet{kotko-lasota2012}    {have also used} the DIM, which is able to 
reproduce the normal outbursts of dwarf novae, {in order to relate the amplitudes of the outbursts
and the} recurrence times. Both methods yield $\ahot \sim 0.1-0.2 $.

Similarly, to reproduce the light curve of \src~in the model of the cooling 
front propagation without irradiation, we adopt the approximations to 
numerical modelling by \citet{menou_etal1999fronts}. The accretion rate decays 
as
\begin{equation}
  \dot M (t)  = \dot M_\mathrm{peak} \, 
(R_\mathrm{front}(t)/R_\mathrm{hot,peak})^{2.2}\, ,
 \label{eq.mdot_cf}
 \end{equation}
where the radius of the hot zone
$$
R_\mathrm{front}(t) = R_\mathrm{hot,peak} -  u_\mathrm{front}\,t
$$
can be found using the front velocity
\begin{equation}
 u_\mathrm{front} =  k\, \alpha \, u_\mathrm{sound}\, , \quad u_\mathrm{sound} 
= \sqrt{\Re\, T_\mathrm{crit}/\mu}\,
 \, , \quad k\approx 1/14\, .
 \label{eq.cf_velocity}
 \end{equation}
 {At the front, the disk  temperature  at the central plane is} $T_\mathrm{crit}  = 4.7 \times 
10^4$~K~\citep{kotko-lasota2012} and  the molecular weight $\mu = 0.6$. Since irradiation is not 
important,  the maximum  radius of the hot disk is found from the rate of the 
viscous heating \eqref{eq.qvis}  at the peak:
  \begin{equation}
  Q_\vis (R_\mathrm{hot,peak}) = \sigmaSB\, \Thot^4 \, , \qquad  
\Thot=10^4~\mathrm{K}\, .
  \label{eq.rhot_qvis} 
  \end{equation}
  
  As can be seen from Fig.~\ref{fig.Cirr_ak}, the above approach is 
appropriate for \src~(2002) if $C_\irr \lesssim 1.5\times 10^{-4}$. The low set 
of symbols in Fig.~\ref{fig.Cirr_ak}  represents the critical values of {the}
irradiation parameter $C_\irr$ for \src~in 2002, below which the 
self-irradiation could not control the disk evolution. The values are 
obtained from the condition that $Q_\irr/Q_\vis (\rhot) =1  $  at the burst 
peak, taking into account Eq.~\eqref{eq.rhot}.  {Since} ratio \eqref{eq.irr_vis} 
decreases    with the radius, {the effect} of irradiation becomes only 
less during the decay as $\rhot$ decreases. 
  
{By applying} the approximation of 
the {constant-speed} cooling front, $\alpha$ {may} be expressed as 
follows:
\begin{equation}
  \ahot \approx 0.16 \,\frac{\rhot^\mathrm{max}}{\Rsun} \,
   \left(\frac{t_\mathrm{exp}}{10^\di}\right)^{-1}\,
   \left(\frac{k}{1/14}\right)^{-1}\,
   \left(\frac{u_\mathrm{sound}}{25\, \mbox{km\,s}^{-1}}\right)^{-1}\, .
   \label{eq.ahot_cf1}
\end{equation}

Expressing the hot-zone size at the peak $\rhot^{\mathrm{max}}$ from the  the accretion rate $\dot 
M_\mathrm{max}$ {using} \eqref{eq.rhot_qvis}, we rewrite the last relation as:
\begin{equation}
  \ahot \approx 
  0.07
   \left(\frac{m_\x \,\dot 
M_\mathrm{max}}{10^{18}\,\mathrm{g\,s}^{-1}}\right)^{1/3}
    \left(\frac{10^\di}{t_\mathrm{exp}}\right)
   \left(\frac{10^4\,\mbox{K}}{T_\mathrm{hot}}\right)^{4/3}
   \left(\frac{1/14}{k}\right)
   \left(\frac{25\, \mbox{km\,s}^{-1}}{u_\mathrm{sound}}\right) .
   \label{eq.ahot_cf2}
\end{equation}
 
{The numerical} factor in 
\eqref{eq.ahot_cf2} depends on the power-law  index of $\dot M$($\rhot)$ in 
{relationship} \eqref{eq.mdot_cf}.

\section{Results of  fitting of the accretion-rate evolution for \src~(2002)}\label{sec.results}

\subsection{Mild irradiation}

\begin{figure}
\rotatebox{0}{{\resizebox{0.45\textwidth}{!}
{\includegraphics{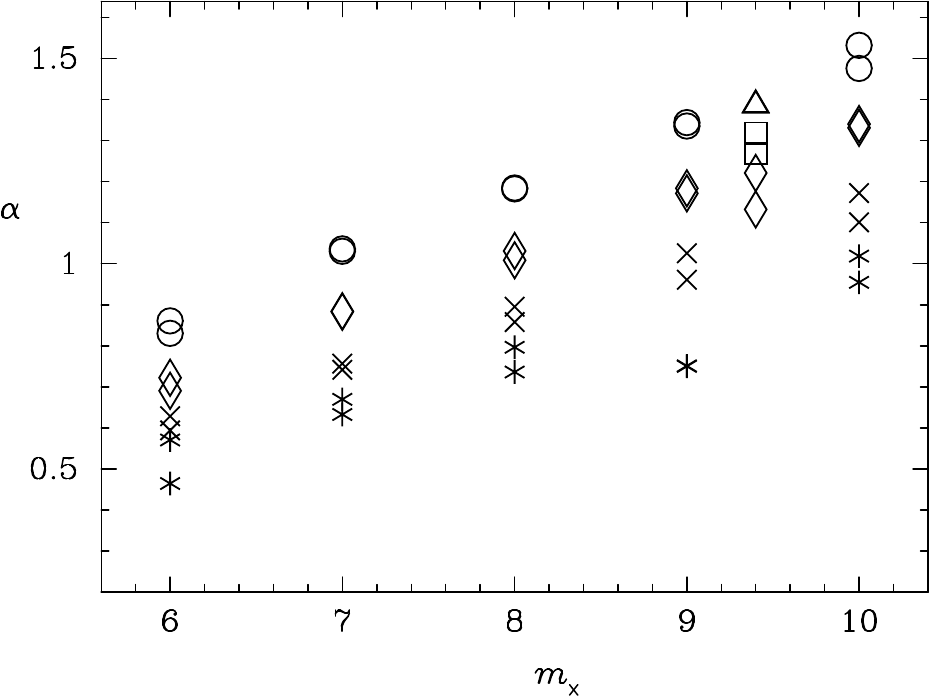}}}} \hskip 0cm
    \caption{{The $\alpha$-parameter, which the ionized disk in \src~(2002) had, if its size was controlled
    by irradiation (\S\ref{s.irradiation_dominated_evolution}) }.  
    Values of $\alpha$ are obtained for values of $C_\irr$ shown in  
Fig.~\ref{fig.Cirr_ak} by bigger symbols. Each value of $\alpha$ is 
obtained {using} $\chi^2$ minimization of the $\dot M(t)$ curve.  {The radius}  of the hot 
zone corresponds to  $T_\irr = 10^4$~K.  As before, the symbols indicate 
{different values} of $\ak$: 0  (circles), 0.1 (triangle), 0.4 (square), 0.6 (diamonds), 0.9 
(crosses), and 0.998 (asterisks). Each pair {of parameters}  ($m_\x$,$\ak$) has two resulting 
$\alpha$ for two {\sc XSPEC} models, {with either} $simple$ or $comptt$ 
component.
   }
\label{fig.alpha-Cirr}
\end{figure}

The viscous evolution of the hot part of the disk  {whose size is}  
controlled by irradiation (\S\ref{s.irradiation_dominated_evolution})  can 
explain the shape of  the observed curve $\dot M(t)$. For characteristic values of $C_\irr$ 
shown in Fig.~\ref{fig.Cirr_ak} (the upper boundary between the empty and shaded 
areas, $C_\irr = (2-3)\times 10^{-4}$), we  {solve equation \eqref{eq.diffusion} numerically} using code {\sc 
freddi} to find $\dot M(t)$. Comparing it with $\dot M(t)$ derived from  the 
spectral modelling (see \S~\ref{sec.sp_fit}), we  find the best-fit values of 
$\alpha$ for a grid of BH parameters (Fig.~\ref{fig.alpha-Cirr}).

{Higher values of $C_\irr$ would correspond to larger hot-zone size $\rhot$, according to \eqref{eq.rhot}. Since the observed $t_{\exp}$ is fixed, this would require bigger values of $\alpha$.}

\begin{figure}
\rotatebox{0}{{\resizebox{0.46\textwidth}{!}
{\includegraphics{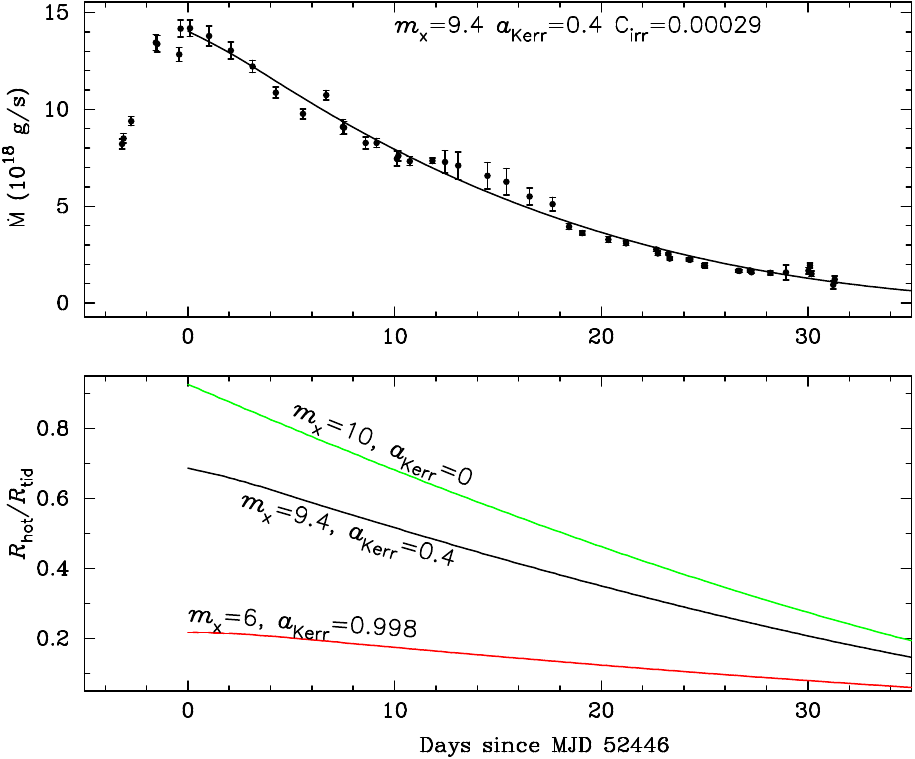}}}
} \hskip 0cm
\caption{{The model of a hot zone controlled by irradiation}. The top panel: modelled  and  
observed $\dot M(t)$ of \src~(2002)  for the set of parameters indicated on the plot. 
The lower panel: {the evolution of the hot-zone radius} $\rhot$  in units of the tidal radius $R_\mathrm{tid}$:
{the } black curve is the model from the top panel; {the} green and red curves are extreme cases with the parameters 
indicated.
{The tidal} radius $R_\mathrm{tid}$ depends on $m_\x$ (see Fig.~\ref{fig.Rpeak1}).
}
\label{fig.Cirr_star_opt_evol}
\end{figure}

  An example of a  model  with {the minimal} $\chi^2$ statistic is presented in 
Fig.~\ref{fig.Cirr_star_opt_evol}.  The top panel shows the evolution  of {
the rate of accretion} through the inner radius for the  parameters indicated in the plot, along with the spectral-modelling
results from Fig.~\ref{fig.sp_fit_res}.
In the lower panel of Fig.~\ref{fig.Cirr_star_opt_evol}, the
evolution of the irradiation-controlled hot-zone size is shown 
for the same parameters by black.
     By green and red (online), we show the hot-zone radius evolution for two other sets of parameters, representing extreme 
     cases within the parameters' limits considered. Values of the accretion
     rate and distance to the source are unique for each set.
  

{Vlues of $\alpha$, 
found by {\sc freddi} for such scenario, lie in the interval $\sim 0.5-1.5$}~(see 
Fig.~\ref{fig.alpha-Cirr}). It is possible that {\sc freddi} does not fully 
account for complex effects appearing  when the outer boundary $\rhot$ is 
moving. This can bias the resulting $\alpha$ towards bigger values.
 We discuss {this} further in \S\ref{ss.ld}.
The results, presented in Fig.~\ref{fig.alpha-Cirr}, can be approximated in a way similar
 to \eqref{eq.f_kramers} or \eqref{eq.f_opal}. The approximation has
 the the same power indexes from the disk parameters, while the dimensionless numerical
 factor is less:
 0.17, for Kramers opacity, and 0.15, for OPAL approximation, 
with {an accuracy of} $\sim 12\%$.

 If  $C_\irr$ lied in the interval $(1.5-3) \times 10^{-4}$, {it would follow from the picture} described 
in~\S\ref{s.irradiation_dominated_evolution} and \ref{s.cf}, {that} the cooling front of `dwarf-nova-disk' type had started sometime during the investigated time interval.  A  visible change in 
the evolution of  $\dot M(t)$   is expected, when irradiation fails to support 
the  outer hot disk~\citep{dubus_et2001}, but further modelling is needed to 
study this in detail.  Notice that there are no suspicious turns 
in the $\dot M(t)$ curve during MJD 25446$-$25474 (see the note at the end of \S\ref{sec.sp_fit}).
    
    \subsection{No irradiation}
    
\begin{figure}
\rotatebox{0}{{\resizebox{0.45\textwidth}{!}
{\includegraphics{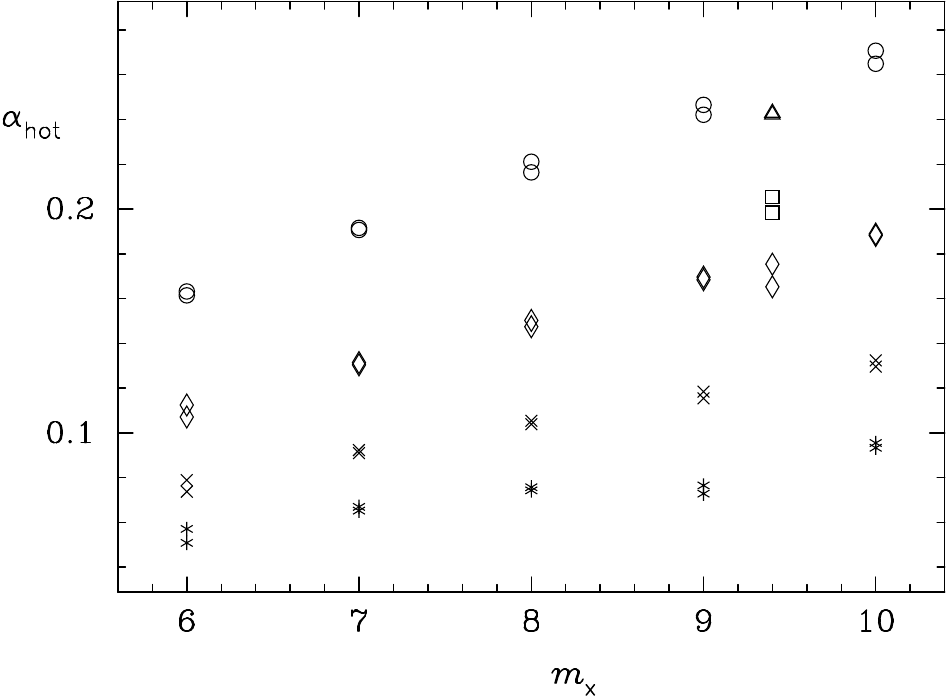}}}} \hskip 0cm
 \caption{{The best-fit} $\alpha$ obtained for the burst of \src~(2002) in the 
approximate cooling-front model~\citep{menou_etal1999fronts}. The accretion 
rate evolves {in accordance with the} expressions in~\S\ref{s.cf}. Different symbols 
indicate 
{values} of $\ak$ as in Fig.~\ref{fig.alpha-Cirr}.}
\label{fig.alpha-Mdot-cf}
\end{figure}

For $C_\irr \lesssim 1.5\times 10^{-4}$, we can describe the burst decay using 
the same approach as used for normal outbursts of  dwarf novae 
(see~\S\ref{s.cf}). For the cooling front moving with {a} constant speed and 
the peak radius of the hot zone corresponding to $T_\mathrm{hot} = 10^4$~K, 
we obtain $\alpha$ shown in  
Fig.~\ref{fig.alpha-Mdot-cf} for 
the same grid of the values $m_\x$ and $\ak$. All values of $\alpha$ are in the interval   $0.08-0.32$.

\begin{figure}
\rotatebox{0}{{\resizebox{0.45\textwidth}{!}
{\includegraphics{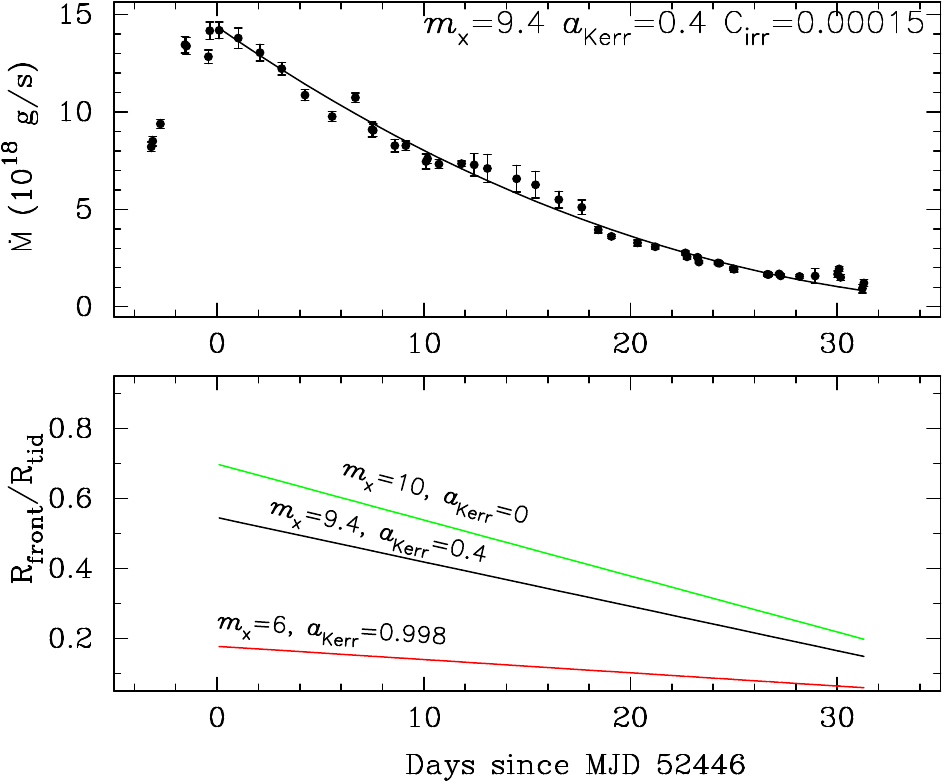}}}} \hskip 0cm
\caption{{The} model of the cooling front propagating with {a }constant speed 
without irradiation.   Top panel: evolution of $\dot M(t)$  for parameters $\ahot=0.23$,  
$m_\x=9.4$, $\ak=0.4$. Lower panel: the radius of the hot zone for these (black) and other parameters (green and red, online).
}
\label{fig.cf_opt_evol}
\end{figure}

Fig.~\ref{fig.cf_opt_evol} shows the modelled 
accretion rate for the specific BH parameters: $m_\x=9.4$ and $\ak=0.4$ (the top panel). 
The lower panel of Fig.~\ref{fig.cf_opt_evol} shows the hot-zone radius evolution for three sets of the parameters.
As can be seen from {the} comparison of  Figs.~\ref{fig.Cirr_star_opt_evol} and \ref{fig.cf_opt_evol}, 
{ the hot-zone radius is less for the no-irradiation scenario for the same disk parameters. The smaller size
of the hot disk is one of the reasons for lesser values of $\alpha$. Notice that the optical observations
could be used to constrain the radius of the hot disk.}

\section{Optical emission from {the } disk in \src~(2002)}\label{s.optical_emission}
\begin{figure}
\rotatebox{0}{{\resizebox{0.48\textwidth}{!}
{\includegraphics{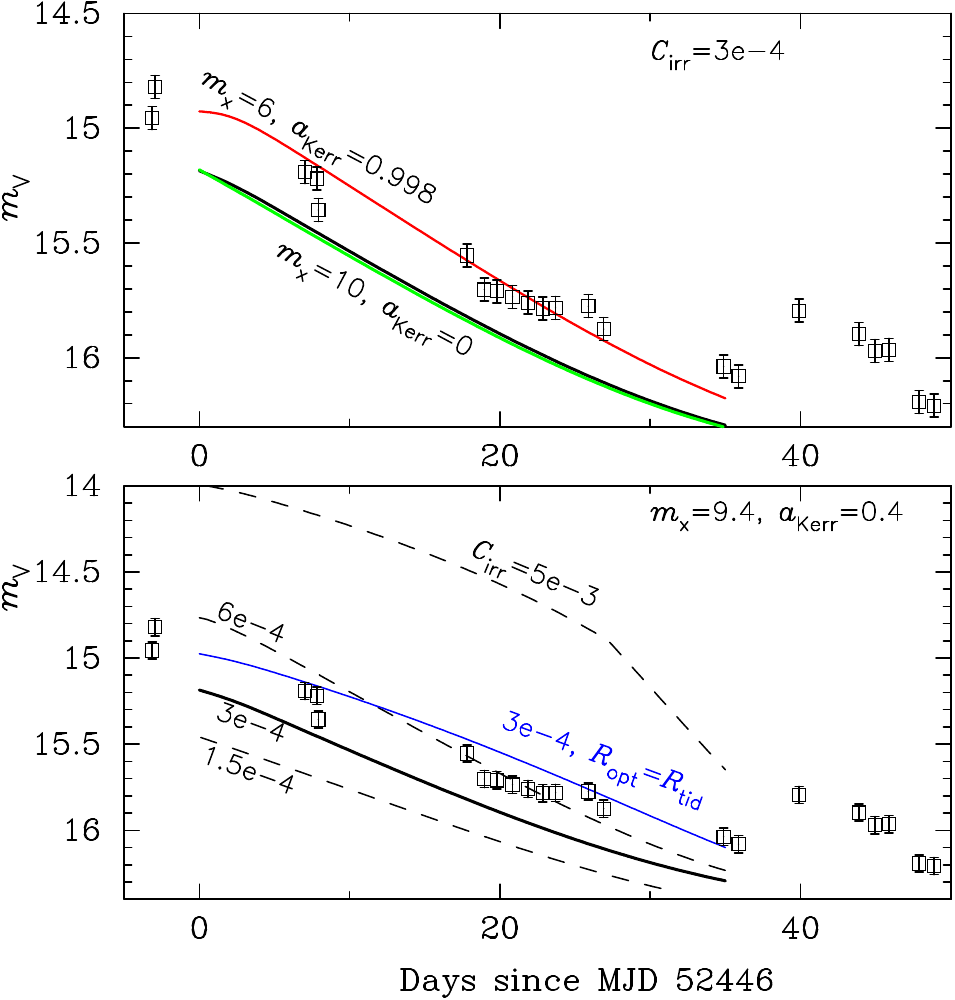}}}} \hskip 0cm
\caption{
Observed and modelled optical $V$ light curves of \src. The model flux is produced by 
the hot zone and calculated using observed $\dot M(t)$. The 
interstellar extinction and pre-burst flux are included. Observational points 
are the ISM-reddened data from \citet{buxton_bailyn2004}. The top panel shows 
models with $C_\irr=3\times 10^{-4}$ and extreme values of the BH parameters, 
indicated near the curves (red and green line). The bold black curve
(close to the lowest green line  in the top  panel) is obtained for $C_\irr=3\times 10^{-4}$, $m_\x=9.4$ and 
$\ak=0.4$. In the lower panel, modelled light curves are presented for fixed BH 
parameters  and different values of $C_\irr$.  The higher solid curve (blue 
online) is the flux produced by the disk with constant size equal to the tidal 
radius.
}
\label{fig.lcV}
\end{figure}
\begin{figure}
\rotatebox{0}{{\resizebox{0.48\textwidth}{!}
{\includegraphics{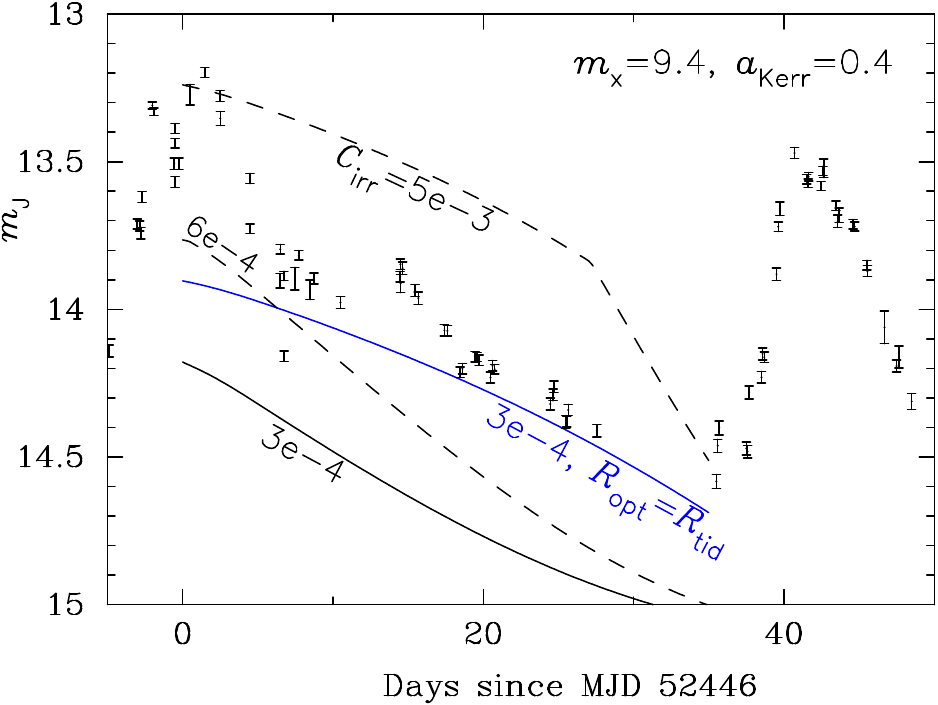}}}} \hskip 0cm
\caption{
Same as the lower panel of Fig.~\ref{fig.lcV} but for the $J$ optical band.}
\label{fig.lcJ}
\end{figure}

Last decade studies {have} provided evidence that the optical emission during an X-ray 
Nova outburst could be produced not only by the outer parts of the disk 
illuminated with the central radiation. The low-frequency portion of the 
non-thermal emission from the BH vicinity can contribute to optical flux. 
Studies of correlation between X-ray, optical, infrared, and radio emission of 
some X-ray transients suggest that jets are responsible for at least a part of 
the emission~\citep[e.g.,][]{russell_etal2006, rahoui_etal2011}. Non-thermal 
electrons in the central disk corona can produce power-law  OIR spectra (the 
hybrid hot-flow model; see \citet{poutanen-veledina2014}). 

{Here we consider 
the optical emission only from the accretion disk.}
Figs.~\ref{fig.lcV} and \ref{fig.lcJ} present {a} comparison of the modelled 
optical light curves with the data {of} \citet{buxton_bailyn2004}.  
If not stated otherwise, the optical flux is integrated  only from the hot part of 
the multi-colour disk using $\dot M(t)$ derived from  the 
spectral modelling. The effective temperature of a disk ring is calculated
as $T_\eff^4 = (Q_\irr + Q_\vis)/\sigma$.  

The modelled optical flux depends on the irradiation parameter $C_\irr$, BH mass  and
Kerr parameter (because the mass and accretion efficiency affect derived $\dot M$),  and 
hot-zone radius.  The hot-zone radius itself depends on $C_\irr$ via \eqref{eq.rhot} or  {on} $\alpha$ via \eqref{eq.cf_velocity}. 
{It is important} to note that {a} value of the turbulent
parameter $\alpha$, derived in the previous section,   is a dynamical characteristic 
of the disk and  depends on $C_\irr$ but does not affect directly the observed optical flux
in the case of the  irradiation-controlled hot zone.

{The  
$V$ and $J$ optical bands} are chosen because there are data in these bands  just before the 
{onset} of the burst. Spectral {flux densities} at $\lambda_V = 5500~\AA$ and 
$\lambda_J = 12600~\AA$ are converted to {the} optical $V$ and $J$ magnitudes using 
the zero-fluxes $F_V^0= 3.750 \times 10^{-9}$~erg~s$^{-1}$~cm$^{-2}$~$\AA^{-1}$ 
and $F_J^0= 3.021 \times 10^{-10}$~erg~s$^{-1}$~cm$^{-2}$~$\AA^{-1}$ 
\citep{cox2015}. {The optical} interstellar extinction  is {accounted for} in the plotted 
light curves {as} $A_V = 1.6$~mag \citep{orosz_etal1998} and $A_J = 
0.282\,A_V$~\citep{buxton_bailyn2004}.

To plot the model curves, we add the modelled optical {fluxes} to the pre-burst 
flux, {namely,} $16.43$ mag in $V$  and $15.13$ mag in $J$. The modelled magnitudes do 
not include {a possible optical flux from other sources: due to} reprocessed X-rays by the 
outer cold disk, companion star, jet, or corona. There is an exception though: 
we show the upper limit on the flux from the  disk, calculated as if the accretion rate
is uniform up to  $R_\mathrm{tid}$: the curves with {the notation} `$R_\mathrm{opt}=R_\mathrm{tid}$' in
Figs.~\ref{fig.lcV} and \ref{fig.lcJ} (blue online).


Emission in  {the} $J$ band  is much  {more} variable than in $V$. 
\citet{buxton_bailyn2004} {have shown} that the optical/IR spectrum at the secondary 
maximum {arizing $\sim 40$~days} after the peak is explained by a jet 
(low-frequency  power-law component in the spectrum). {Other 
large $J$ variations}, visible before {the 40th~day}, {may also be associated with emission 
of this type}.

{{Figures~\ref{fig.lcV} and \ref{fig.lcJ} demonstrate} that the irradiation 
parameter $C_\irr\sim 5\times 10^{-3}$ is excluded, otherwise the optical flux 
from such a disk would greatly exceed the} observed $V$ and $J$ values. From the $V$ data 
alone, $C_\irr\approx(3-6)\times 10^{-4}$ is suggested, taking into account 
possible extra emission from the irradiated outer cold disk. {A value of  $C_\irr$ deduced from the $J$ data is approximately the same, although}
 there is {some} uncertainty due to 
{large} flux variations in the $J$ band.  Very weak irradiation,  $C_\irr\approx1.5\times 10^{-4}$,
 needed to {allow} the  dwarf-nova type evolution (\S~\ref{s.cf}), is less preferred from the 
point of {view of} optical observations but could not be excluded altogether in view of  other possible
sources of the optical photons.

\section{Discussion}\label{sec.disc}

\subsection{Values of $C_\irr$}\label{ss.disc_cirr}

 There are several similar definitions of $C_\irr$ in the literature 
depending on how the central luminosity is {expressed}. \citet{esin_et2000} {have used} the 
total X-ray luminosity obtained from analysing the X-ray data, and {this} 
$C_\irr$ is greater by a factor of $L_\mathrm{tot}/L_\mathrm{X}$ than that defined by \eqref{eq.C_irr_def}; 
\citet{dubus_et1999} {have expressed} the irradiation flux {using} $\dot M\, c^2$, {which decreases}
$C_\irr$ by a factor of the accretion efficiency $\eta$.
In \eqref{eq.C_irr_def} we follow \citet{dubus_et2001} and 
\citet{suleimanov_etal2008}, bearing in mind that $L_\mathrm{tot} $ does not 
depend on {an} observational band or inclination of a binary system. {When trying to check}
 the idea of $C_\irr$ being a universal value for X-ray transients, it's 
preferable to free oneself from additional biases. On the other hand, {a
large degree of uncertainty may be involved with the} factor $C_\irr$ due to unknown geometry factors, 
albedo, and thermalization efficiency.

 The irradiation parameter can be written as 
follows~\citep{suleimanov_et2007e}: 
\begin{equation}
 C_\irr =  (1-A_X)  \, \Psi (\theta) \, \left( \frac{\di z}{\di  
r}-\frac{z}{r}\right)\, ,
 \label{eq.C_irr_expres}
\end{equation}
where $\Psi(\theta)$ is the angular distribution of the central flux, $z$ is 
the height of the interception of the central irradiation, $\theta$ is the 
polar angle of the vector {from the disk centre}
to the intercepting unit surface. The term in the 
brackets {determines} the inclination of the illuminated surface to the incident 
radiation. Factor $1-A_X$, equivalent to {the} `thermalization coefficient',
{determines a} portion of the intercepted bolometric flux that is absorbed and 
reprocessed to the black-body radiation. For the 
Newtonian metric and {a} flat disk, we have $\Psi (\theta)=2\, 
\cos(\theta)\approx 2\, z/r$. {An inclusion} of the general relativity effects 
modifies the angular distribution even for a non-rotating black hole. For an 
extremely fast rotating black hole, the angular distribution becomes remarkably 
flat~\citep[see figure 9 of][]{suleimanov_et2007e}. Here we assume that the 
disk is thin and the distance {from the centre to the illuminated surface} equals 
the cylindrical coordinate $r$.

 Let us consider the case of the direct illumination of the $\alpha$-disk 
surface. For a simple model of a central X-ray source  with the geometry of a 
flat disk around a stellar-mass compact object~\citep{suleimanov_et2007e}: 
\begin{equation}
C_\irr \sim 6 \times 10^{-5}\, \left(\frac{z_0/r}{0.05}\right)^2 \, 
\frac{1-A_X} {0.1}\, .
\label{eq.c_irr_th} 
\end{equation}

 Factor $1-A_X$ depends  on the disk albedo and  spectral-dependent 
efficiency of reprocessing X-rays into thermal photons. The shape of {an} X-ray 
spectrum is important: mainly the relatively hard X-rays ($>3$~keV) penetrate deep enough 
into the disk to be thermalized efficiently, and $1-A_X \sim 
0.05-0.1$~\citep{suleimanov_et1999}. The formula above does not  take into 
account the relativistic amplification due to the photon focusing. This 
amplification factor can reach $3-4$ for a rapidly rotating Kerr black hole 
with $\ak=0.9981$ in  the  direction $\cos(\theta)= 
0.1$~\citep{suleimanov_et2007e}.

 Formula \eqref{eq.c_irr_th} demonstrates that generally $C_\irr$ is 
variable: {the} thermalization coefficient $(1-A_X)$ changes {if} the X-ray 
spectrum {becomes softer}, and the half-height $z_0$ of the disk gets smaller with 
decreasing accretion rate. Moreover, the magnitude of $C_\irr$ in 
\eqref{eq.c_irr_th} {does not seem to be large enough to be able}  to explain the optical data of X-ray 
transients, even with relativistic effects taken into 
account~\citep{suleimanov_et2007e}. {Different ideas 
have been put forward to eliminate this discrepancy}. For example, illumination through {an inhomogeneous corona} above the disk  {may yield a}  higher  factor 
$C_\irr$~\citep{suleim_et2003,suleimanov_et2007e, gierlinski_etal2009}. 
\citet{suleimanov_etal2008} and \citet{mescheryakov-rev-fil2011} {have shown} for some 
LMXBs that the effective geometric thickness can be twice {as large as} the hydrostatic 
thickness of the $\alpha$-disk. On the other hand, {according to} \citet{dubus_et1999}, 
the cold disk {may be shadowed by the hot disk from 
the  central  irradiation} and, thus, factor $C_\irr$ in the cold 
disk {might}  be much smaller.

 The haracteristic 
values of $C_\irr$  (Fig.~\ref{fig.Cirr_ak}), obtained for \src~(2002), are  smaller 
comparing to ones {usually derived on the theoretical and observational grounds}. 
The irradiation factor $C_\irr$ {should be}  $ < 6\times 10^{-4}$ 
{in order for the hot-zone radius to be variable}.  {On the other hand, 
\citet{dubus_et2001} adopted} $C_\irr\sim 5\times 10^{-3}$ 
to {reproduce} the light curves of X-ray novae.  {When analysing} {\em Swift} 
optical/UV/X-ray broad-band data of XTE\,J1817$-$330 (the outburst  {of} 2006), 
\citet{gierlinski_etal2009}  {have found out} that the spectra are consistent with a 
 {model of} reprocessing  {a constant fraction of} $10^{-3}$ of the bolometric X-ray 
luminosity (disk plus non-thermal tail),  {arguing} in favour of  {the} direct 
illumination of the black hole disk. For the 1999-2000 outburst of  {the} BH 
transient XTE\,J1859+226, \citet{hynes_etal2002} have estimated  {the} irradiation 
parameter $C_\irr \sim 7.4\times 10^{-3}$, assuming  {the} accretion efficiency  {to be}
$\eta_\mathrm{accr} =0.1$.

Estimate \eqref{eq.c_irr_th} agrees with {the} value of $C_\irr$ for \src,  {assuming a presence of}
some relativistic focusing and the fact that the relative 
half-thickness $z_0/r$ near $\rhot$ at the  peak reaches $\sim 0.07$. There is no need to 
contrive {ways for increasing} the value given by \eqref{eq.c_irr_th}. {Instead,} one has to answer a question what 
is the reason for $C_\irr$ being lower { during the  
particular burst  of this X-ray transient.}

{Figure} ~\ref{fig.dotMpeak} demonstrates that the disk was close to the Eddington limit {in the 2002 outburst}.  Above the Eddington limit, outflows from {the} disk are expected 
\citep{shakura-sunyaev1973}. {There was possibly a weak outflow in \src~(2002), which was able to} 
effectively attenuate X-rays. In principle, this {may} 
constrain $C_\irr$ in X-ray transients with about-Eddington accretion rates.

It is also possible   that {the} factor $C_\irr$ depends  on the proximity of $\rhot$ 
to the tidal truncation radius of the disk. In short-period X-ray novae, 
$\rhot$ at the peak is close to $R_\mathrm{tid}$, while {its peak value in \src~$\rhot$ is 
apparently only about $ (0.5-0.7)\, R_\mathrm{tid}$}.

\subsection{Models with irradiation-controlled {size of the hot zone} }\label{ss.ld}
\begin{figure}
\rotatebox{0}{{\resizebox{0.46\textwidth}{!}
{\includegraphics{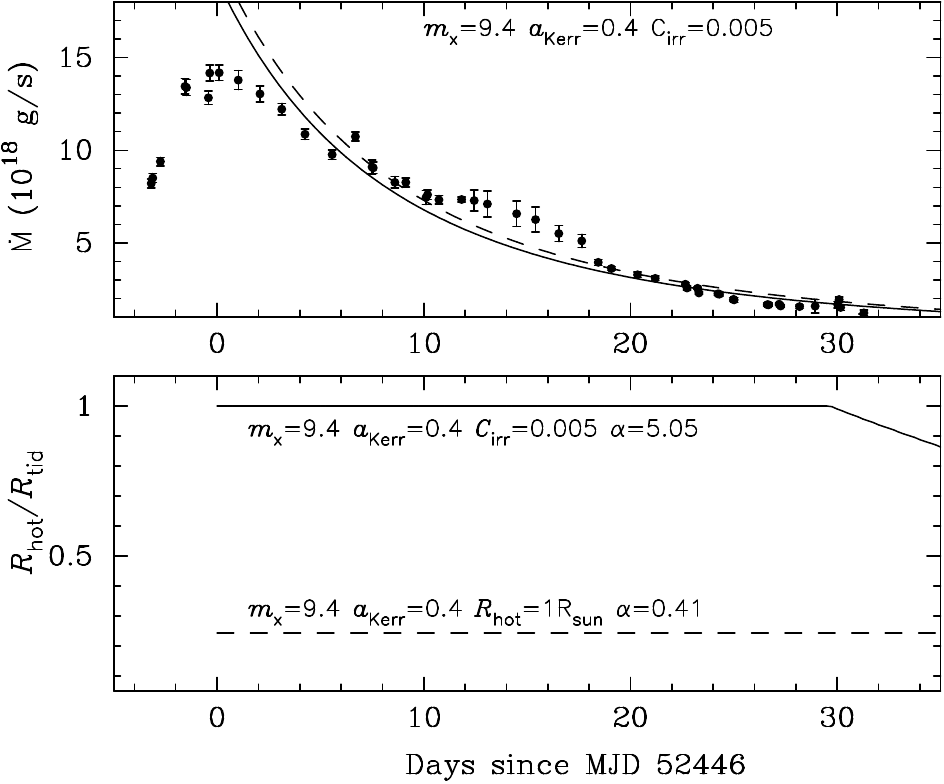}}}}
 \hskip 0cm
\caption{The evolution of a strongly-{irradiated} disk, $C_\irr = 5 \times 10^{-3}$
(solid line), and {evolution} of a disk with {the}  fixed radius of the hot zone (dashed line).  Top panel:
{the } accretion rate found by minimizing $\chi^2$, lower panel: {the}  radius of the hot zone.
}
\label{fig.Cirr_max_opt_evol}
\end{figure}

 {Figure}~\ref{fig.Cirr_max_opt_evol} presents  two 
attempts {of modelling the decay of the \src~ burst in 2002 in the assumption that the radius of the evolving  
disk remained constant}. The plot illustrates that the {assumption of constant size for the hot zone cannot in principle yield the observed shape of $\dot M(t)$}, regardless of the
{values} of $\rhot$ and $\alpha$. 

 The stage of  irradiation-controlled  hot zone with changing size
 was considered by \citet{king-ritter1998}.  The equation for the shrinking hot-zone mass, 
 \begin{equation}
 \dot M_\mathrm{hot} = - \dot M_\mathrm{in} + \frac{\mathrm{d}}{\mathrm{d}t} 
\left(\Sigma(\rhot) \, \piup \, \rhot^2 \right)\,,
 \label{eq.shrinking_zone}
 \end{equation}
 can be solved if one assumes {a quasi-stationary evolution of the disk.}
 In this case, $M_\mathrm{hot} \propto  \Sigma(\rhot) \rhot^2$.   
 {The relationship} \eqref{eq.Sigma_hF} between 
$\Sigma$ and $F$ and condition \eqref{eq.rhot} 
{for determining the radius $\rhot$ of the hot zone yield the following  power-law relationship 
between the mass of the hot part of the disk $M_\mathrm{hot}$ and the inner accretion rate $\dot M_\mathrm{in}$:}
  \begin{equation}
   \dot M_\mathrm{in} \propto M_\mathrm{hot}^{\,40/53}
   \label{eq.mmprop}
  \end{equation}
 {(Kramers opacity). Expression~\eqref{eq.mmprop} is fully confirmed  by the {\sc freddi} calculations. }
  {It can be deduced from \eqref{eq.mmprop}} that $\dot M \propto  (t_\mathrm{end}-t)^{40/13}$, where 
$t_\mathrm{end}$ {stands for the moment in time} when the formal solution gives {the } zero disk mass.
   { Notice that the consideration of this stage 
   for the case,  
    when the kinematic turbulent viscosity $\nut$ is  independent of time and radius,
    yields $\dot M \propto  (t_\mathrm{end}-t)$ \citep[`linear decay'; by][]{king-ritter1998}.
   However, variable $\nut$ is inherent for  $\alpha$-disks.}
  
%


More involved numerical models {used for} calculating  thermal stability and 
evolution of {a} disk with {the} irradiation-controlled cooling front~\citep[see, 
e.g.,][]{dubus_et2001} suggest that {the} simple model expressed by 
\eqref{eq.shrinking_zone}  may not provide {fully reliable $\alpha$ values}. 
{The} quasi-stationary {approximation}  {may prove to be} too crude to {
accurately assess} the velocity of {the} boundary of the uniform-viscosity zone.

To summarize, the model of the irradiation-controlled shrinking hot zone implemented in  {\sc 
freddi} describes the shape of $\dot M(t)$ in \src~quite satisfactorily. On 
the other hand, its current version probably overestimates $\alpha$.  {Notice that the 
model of a completely ionized disk, also calculated by {\sc 
freddi}, does not suffer from this shortcoming.}

\subsection{Determination of $\dot M$ from observations }

{ It is important to consider} the general relativity effects {when analysing the} 
disk evolution in real systems. For given X-ray data, the  peak 
accretion rate depends strongly on  the Kerr parameter $\ak$: $\dot M$ changes 
by a factor of $\sim 25$ for $\ak=0-0.998$. This factor incorporates  the 
variation of accretion efficiency of rotating black holes with different $\ak$. 
Also, in the vicinity of a black hole, the outgoing X-ray spectrum is disturbed 
by the effects of Doppler boosting, gravitational focusing, and the 
gravitational redshift~\citep{cunnin1975}. 

 In the present study, we {use} the spectral model of the relativistic disk $kerrbb$ developed by  \citet{li_et2005}. 
{As mentioned by the authors,} $\dot M$ in $kerrbb$ is the `effective' accretion 
rate, while an actual one should be greater by a factor of {$(1+\eta_\mathrm{in.t.})$ }, where 
$\eta_\mathrm{in.t.}$ is  the ratio of the disk heating {due to} a non-zero 
inner torque to {the heating caused by} the fall of accreting matter. {For} a 
non-rotating black hole, the magnetic torque at the inner edge of {a thin
magnetized} accretion disk is only $\sim 2\%$ of the inward flux of the angular 
momentum~\citep{shafee_et2008}. In the present work, we assume  
$\eta_\mathrm{in.t.}=0$. 

If the innermost stable circular orbit {is} $0.5 \,G\,M_\x/c^2$ closer {to the centre} than a 
canonical value   {of} $6\, GM_\x/c^2$ for a non-rotating black hole,  as argued by \citet{reynolds-fabian2008}, {then the} 
related {disk-power}  increase corresponds to {a}  formal  variation of  $\ak$ from 0 
to 0.15. As can be seen from  Fig.~\ref{fig.dotMpeak}, {a } higher Kerr parameter {for}
the black hole translates into {a} smaller  accretion rate.  This uncertainty of 
relativistic disk models tends to be smaller for higher 
$\ak$~\citep{miller_et2009}. {In view} of such uncertainties, we {varied
 the Kerr parameters within some range} to see how the results  depended on them.

There are other  modern {\sc XSPEC} models for emission of the relativistic 
disk, for example,  $kerrbb2$ and $slimbh$, which are more sophisticated in 
some respects. {However, these}  have {the} limitations  that hinder us from 
using them: {the $kerrbb2$ model} covers limited {ranges for the accretion rate and spectral energy};  $slimbh$ has {the} luminosity as an input parameter, instead of the accretion 
rate.

\subsection{Variation of the model assumptions}\label{ss.variations}

\begin{figure}
\rotatebox{0}{{\resizebox{0.45\textwidth}{!}
{\includegraphics{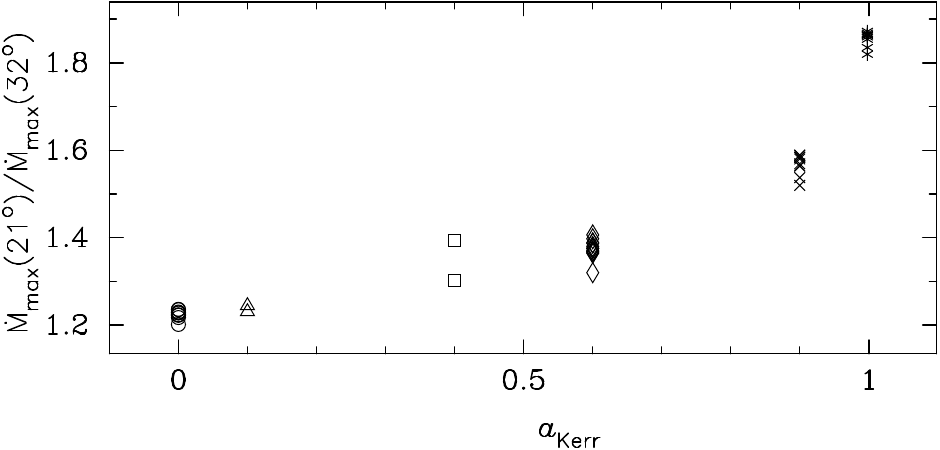}}}} \hskip 0cm
\caption{{The ratio} of the peak accretion {rates of the  \src~(2002) outburst}
 for $i=21\degr$ and $i=32\degr$.
 {The designations} are {the same} as in Fig.~\ref{fig.dotMpeak}.  {The dependence} on $\ak$ is due to the
 relativistic focusing of X-rays.}
\label{fig.dotMpeak32}
\end{figure}

Results of spectral fitting by \citet{morningstar-miller2014} suggest that the 
inclination of the inner disk in \src~ {differs} from the binary 
inclination. {When taking as the inclination the central value from the range} derived by 
\citet{morningstar-miller2014}, $i=32\degr$, we obtain {lower resulting accretion 
rates}. The decrease depends on $\ak$: 
it is $\sim 86\%$ for $\ak = 0.998$ and $\sim 24\%$ for $\ak=0$~(Fig.~\ref{fig.dotMpeak32}). {This} 
inclination makes {the} resulting $\alpha$ smaller by {a} factor {of}  1.3 for $a=0.998$ and by 
1.1 for $\ak=0.0$ (for the Kramers opacity), as can be 
 found from {relationship}  \eqref{eq.f_kramers}.

Spectral modelling with the colour factor $\fc=1$ {gives systematically} higher 
peak accretion rates and, thus, higher estimates for $\alpha$.

%

\begin{figure}
\rotatebox{0}{{\resizebox{0.45\textwidth}{!}
{\includegraphics{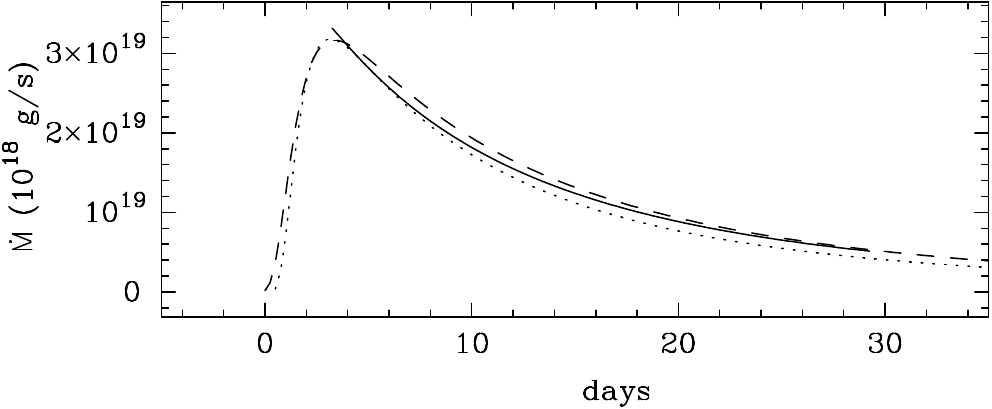}}}} \hskip 0cm
\caption{{The modelled evolution of} $\dot M(t)$  
{with} different opacity implementations for the solar {abundances}: 
the dashed line is for Kramers law $\varkappa = 5\times 
10^{24}\rho/T^{7/2}$~cm$^2$\,g$^{-1}$;
the dotted line is for the analytic approximation to {the} OPAL tables $\varkappa = 
1.5\times 10^{20}\rho/T^{5/2}$~cm$^2$\,g$^{-1}$ 
\citep{bell-lin1994};
the solid line is  calculated {using} the code of \citet{malanchev-shakura2015} 
for {the} OPAL tables~\citep{iglesias-rogers1996}.
Disk radius is constant, $C_\irr=0$.}
\label{fig.comp_opac}
\end{figure}

In Fig.~\ref{fig.comp_opac}, we compare models with three different opacity 
implementations. {The} parameters of the disk are set {as follows}: $m_\x=10$, $\alpha=0.5$, 
$\rhot=10^{11}\,\mbox{cm}=\const$. We see that the difference  {in the opacity} 
{does not affect the results significantly in contrast} to other uncertainties involved in the 
modelling.

\section{Summary}\label{sec.summary}

Bursts of X-ray novae are {crucial}  laboratories to probe models of the 
non-stationary disk accretion. {It is important therefore that  a model of the}
viscous disk evolution takes into account {the self-irradiation 
of the disk  in a self-consistent way}. 

{We assume the concept of the high turbulent parameter $\alpha$ in the hot ionized 
part of the disk and low $\alpha$ in the cold neutral zone. \citet{coleman_et2016} have 
considered a dwarf-nova disk  with $\alpha\sim 0.01$ in the ionized and neutral zones and with
$\alpha$ one order higher  in the zone with the partial ionization, following some recent 
results of  numerical MHD simulations  that have achieved a value of $\alpha \sim 0.1$ for 
the regions with the partial ionization and convection \citep{hirose_et2014}. However, 
such low-$\alpha$ disks  may face a difficulty at  explaining the characteristic times of 
the exponential decays (several tens of days) and  the time delays between the optical and 
X-ray band during outburst rises (several days), observed in some X-ray novae.}

If an  X-ray outburst is produced {due to the} variations of the central 
accretion rate in a completely ionized disk, a nearly exponential decay is 
expected for the disk with {a} constant outer radius.  For such decays,
{which may take place in  short-periodic} X-ray novae, the
turbulent parameter $\alpha$ can be estimated using analytic {relationships} 
\eqref{eq.f_kramers}--\eqref{eq.f_opal} and the disk evolution can be 
calculated using {the {\sc freddi} code}. Slower decays happen in the accreting 
viscous {flows, which are radially expanding} ({ordinarily, this is not a case of a close binary 
system}), or if {a companion star provides additional matter to the disk}. Faster 
decays are due to the progressive shrinking of the viscously-evolving zone (the 
zone with high $\alpha$). Detailed modelling is needed to discriminate between 
these cases because the apparent similarity to an exponential decay may mask 
different scenarios.

{The} difference between the disk instability models {proposed for explaining the bursts} in dwarf and X-ray 
novae is thought to be {mainly due to the major role of} X-ray irradiation in the 
latter group {of objects}~\citep[e.g.,][]{paradijs-verbunt1984}. {The typical} irradiation 
parameter $C_\irr \sim 5 \times 10^{-3}$ was suggested~\citep{dubus_et1999, 
lasota_etal2008}.

We derived and analysed the  accretion rate evolution $\dot M(t)$ during {the first} 30 
days after the peak of 2002 outburst of \src, when it was in {the} high/soft state.
{We found} find that {the observed evolution requires $\alpha \approx 3.2-3.3$  for the irradiation parameter $C_\irr \sim 5\times 10^{-3}$}. This happens because {the } strong illumination leads  to a large size of the hot part 
of the disk, and the short decay time can be {obtained} only with {a} high viscosity 
parameter.
 The shape of the $\dot M(t)$ curve of \src~(2002) cannot be produced by
{a} hot zone with constant size; {rather, it} is consistent with the model {of}
 shrinking high-viscosity zone. 

{{Putting aside} mixed scenarios, there {were} two options {depending on whether or not 
 the central irradiating flux controlled the hot zone size}. {These} corresponded to
$C_\irr  \approx  (3-6)\times 10^{-4}$ (`mild irradiation' case)  and $C_\irr  \lesssim  1.5\times 10^{-4}$ (`no irradiation' case) for \src~(2002).
  In the quasi-stationary model of the  irradiation--controlled hot zone,  we obtained
$\alpha \sim 0.5-1.5$ for the range of the BH masses $6-10~\Msun$.} The  {second case
would resemble}  a decay in a dwarf nova disk and {required} $\ahot\sim  0.08-0.32$.

 The $V$ and $J$ light curves  favour {the} value $C_\irr$ in the range $ 
\sim (3-6)\times 10^{-4}$, {making the model of irradiation-controlled hot zone 
 in \src~more plausible}.  Nevertheless, {there remains the possibility}
 that the X-ray illumination can be of {low} importance for  this and some other
X-ray nova outbursts.

\section*{Acknowledgements}
The authors are grateful to the anonymous referee for suggestions
that have improved the clarity of the manuscript. The work is supported by the Russian Science Foundation grant 14-12-00146.
Authors used the equipment funded by M.V.~Lomonosov Moscow State University 
Program of Development.

\appendix
\section{Spectral modelling}\label{a.sp_mod}
We  analyse the archival data for {the 2002} outburst of \src~obtained  with the 
Proportional Counter Array aboard
the {\em RXTE} observatory. We {have selected} the same  data as
in \citet{park_etal2004}. In particular, long  observations
of 17, 19, 20, and  28 June 2002 were divided in two equal intervals. The 
``Standard-2'' data from all Xe layers of PCU-2 {have been} used.
\begin{table}
\caption{Summary of spectral  parameters used for two  models:
$TBabs*((simpl*kerrbb+laor)*smedge)$ or
 $TBabs*(compTT+kerrbb+laor)*smedge$. {The results} obtained with 
 values in brackets are discussed only in \S~\ref{sec.disc}. 
}
\label{tab.parameters}
\begin{tabular}{crl}\hline
Model  & Parameter & Value \\
component &		&\\
\hline
$TBabs$ 	& $n_\mathrm{H}$ 	& $4 \times 10^{21}$cm$^{-2}$ \\
$kerrbb$	& $\eta_\mathrm{in.t.} = F_\in\,\omega_\in/\eta_\mathrm{accr}\, 
\dot M\, c^2$		& 0				\\
	      &	$\ak$			&  0, 0.6, 0.9, 0.998 \\
	      &	$i$			& $20\fdg7$	~($32\degr$)\\
	      & $m_\x$ 			&		6, 7, 8, 9, 10\\
	      & $\dot M$		& $0-10^{21}$\,\grams \\
	      & $d$			& $0-10^4$~kpc \\
	     & $\fc$			& 1.7 ~(1.0) \\
	     & irradiation flag		&  1 \\
	     & limb-darkening flag	& 1 \\
	     & normalization		& 1 \\
$simpl$		& $\Gamma$	     	&	$0-4.5$ \\
		& $f_\mathrm{SC}$	& $10^{-3} - 0.3$ \\
		& only up-scattering flag & 1 \\
$compTT$	& redshift 		& 0 \\
		& $T0$			& $10^{-4} - 4$~keV \\
		&  $kT$			& $10- 10^3$~keV \\
		& $\tau_\mathrm{plasma}$&  $10^{-2}-200$ \\
		& Geometry switch	& disk \\
		& normalization		& $0 -10^{24}$~photons  cm$^{–2}$s$^{-1}$  \\
$laor$		& $E_\mathrm{line}$     & $5-7$~keV \\ 
		& Index			&	3 \\
		& $R_\mathrm{in}$	&	$1.235-400\, GM_\x/c^2$\\
		& $R_\mathrm{out} $	&	$400\,GM_\x/c^2$\\
		& $i$			& $20\fdg7$	~($32\degr$)\\
		&  normalization	& $0 -10^{24}$~photons 
cm$^{–2}$s$^{-1}$  \\ 
$smedge$	& $E_\mathrm{edge}$     & $6-10$~keV\\  
		& $\tau_\mathrm{max}$	& $0-10$	\\
		&Index			& –2.67	\\
		& Width			&7~keV	\\
  \hline
\end{tabular}
\end{table}

The basic reduction is made with {the} help of the {\tt rex} script of {\sc 
HEASoft}~6.18. The  background PCA model for bright sources is {used}. The good time intervals are selected according to 
the conservative options: {the source 
elevation should have been} greater than 10 degrees; the pointing offset, less than 0.02
degrees;  30 min should {have elapsed} after the observatory {passage through} the South Atlantic 
Anomaly. We calculate dead time corrections for the source and background to 
adjust the exposures, make response matrices, and extract spectra using tools 
{\tt pcadeadcalc2} and {\tt pcaextspect2}.
	
{The spectral} fitting  is done {using} {\sc XSPEC}~12.9.0. Systematic errors of 1\% are 
added via {the} command $system$. {We use the absorption} model $tbabs$  with the hydrogen column 
$4\times 10^{21}$~cm$^{-2}$ (according to the LAB~\citep{kalberla_etal2005} and 
GASS~\citep{kalberla-haud2015} surveys;  see AIfA Hi Surveys tool at 
Argelander-Institut f\"ur Astronomie) {together with $wilm$ abundances}. {The spectral}
fitting is carried out in  the 2.9--25~keV interval.

{To describe the thermal emission from the disk,  {the} $kerrbb$  spectral model is 
used. The spectral parameters can be found in  Table~\ref{tab.parameters}. {The 
 accretion rate $\dot M$ and the distance $d$ are set free}. We fix  the BH 
mass $M_\x$ and {the}  dimensionless Kerr parameter $\ak$. The disk inclination is 
fixed {and equal} to $20\fdg7$~\citep{orosz2003} or, alternatively, to 
$32\degr$~\citep{morningstar-miller2014}, and the flux normalization is set to  {be}
unity. Flags for the disk irradiation and limb darkening are set on, and  {the} two 
values of {the colour factor} $\fc$ are tried: 1.7 and 1. Another  fixed parameter is 
the zero-torque condition at the inner edge ($\eta_\mathrm{in.t.} = 0 $).

To describe the non-thermal component as the emission comptonized in the high-temperature plasma near the disk, 
we use the convolution model  $simpl$~\citep{steiner_et2009}. This empirical model, in which a fraction {of}
seed photons produces the power-law component, can be 
used for any spectrum of the seed photons. The model has  two free parameters: the 
photon power law index and the fraction of scattered photons. {Since} $simpl$ 
redistributes input photons to higher and lower energies,  the energy interval
should be extended to adequately cover the relevant band. Following 
\citet{steiner_et2009}, we compute the model over 1000 logarithmically spaced 
energy bins between 0.05 and 50~keV. We hold the power-law index  $\Gamma$ 
between 0 and 4  and choose the `up-scattering' mode.

As an alternative, we also model the power-law tail with 
$comptt$~\citep{titarchuk1994}, which calculates  the Compton scattering as a 
convolution using the scattering Green’s function. 
In  $comptt$, we set free the seed {photons'}  and plasma {temperatures}, {the } optical 
depth, and {the normalization parameter, choosing the `disk' geometry}.

Following \citet{park_etal2004}, we include a spectral component to fit  a broad 
edge-like absorption spectral feature. It is suggested that the feature near 
7~keV is produced by the K-shell absorption of iron, smeared due to reflection 
or partial absorption of X-rays by the optically thick accretion 
disk~\citep{ebisawa_et1994}. The $smedge$ spectral model  reproduces the 
broadened Fe-absorption line (K-absorption structure) and should be 
multiplied with the continuum spectrum. We confine {the} parameter $E_\mathrm{edge}$, {which 
approximately corresponds} to the energy of the iron 
K-edge~\citep{ebisawa1991}, {to values within} 6 to 10~keV, and the smearing width parameter 
is fixed to {be} 7~keV, as in  \citet{park_etal2004}.

The spectral model $tbabs * (simpl*kerrbb)*smedge$  gives acceptable fits 
({the} reduced $\chi^2<1.5$) for the most spectra {with chosen masses and Kerr parameters .}

\citet{park_etal2004} have found an evidence of the Fe $K\alpha$-line in the 
spectra of the source during the outburst. They {have concluded} that the {\sc XSPEC}  {$laor$
model that takes   into account the relativistic effects leading to} line 
broadening  \citep{laor1991} suits the line component better than the Gaussian 
model. We confirm this conclusion. In $laor$, the  {energy of the  line centre is 
 varied freely} between 6.4 and 7~keV, following \citet{park_etal2004},  while the inner radius 
is thawed. {The other}  parameters, frozen by default, remain {unchanged.}
%

\citet{morningstar-miller2014} {have used the} spectral component $relconv${$*$}$reflionx$ 
to describe the feature near 7~keV. We {could not use it} because 
{the spectral component}  {is not able to provide} satisfactory results  at times close to the peak, 
apparently due to {a} limitation on the photon index in $reflionx$. 
 
\section{{THE} {\sc freddi} code overview}\label{a.code}
{The} {\sc freddi}\footnote{{\sc freddi} -- Fast Rise Exponential Decay: accretion 
Disk model Implementation. The code can be downloaded
from \url{http://xray.sai.msu.ru/~malanchev/freddi/}} code is designed to solve 
the differential equation~\eqref{eq.diffusion} 
with two boundary conditions on the  viscous torque~$F$: $F_\mathrm{in} = 0$ 
and $(\partial F / \partial h)|_\mathrm{out} = \dot M_\mathrm{out} = 0$.

 The code uses the analytic {relationship} between the surface density $\Sigma$ 
and {the} viscous torque $F$:
\begin{equation}
    \Sigma = \frac{ (G\, M_\x)^2\, F^{1-m} }{ 4\,\piup\, (1-m)\, D \,h^{3-n} 
}\, ,
\label{eq.Sigma_hF}
\end{equation}
where $m$ and $n$ are the dimensionless constants that equal $3/10$ and $4/5$, {respectively}, 
for  the Kramers opacity law, and  $1/3$ and $1$, for $\varkappa \propto \rho / 
T^{5/2}$~\citep{bell-lin1994}, $D$ is a diffusion 
coefficient~\citep{lyub-shak1987,suleimanov_etal2008}. The dimensional diffusion 
coefficient $D$ in \eqref{eq.Sigma_hF} is approximately constant for specific 
disk parameters, as it depends  on the  parameter $\tau_0$
(comparable to the disk optical thickness), and the {dependence} is rather weak 
for $\tau_0 \gg 1$ \citep{suleimanov_et2007e}. We use {a} constant value of $D$ 
corresponding to $\tau_0 = 10^3$.

In {\sc freddi}, the outer radius {$\rhot$ of the evolving disk can be varied} so 
that either the effective temperature $T_\mathrm{eff}$ {or the} irradiation temperature 
$T_\irr$ {would be} constant. Thus, the {shift} of $\rhot$  tracks the temperature variation. The 
outer boundary condition $\dot M_\mathrm{out} = 0$ means that  
the distribution of the viscous torque in the outer cold disk is ignored.

{To solve} diffusion equation~\eqref{eq.diffusion}, {it is necessary to know} an initial 
distribution of {either the} function $F(h)$ or {the function} $\Sigma(h)$, where $h=\sqrt{G\,M_\x\,r}$ is 
the specific angular momentum. {\sc freddi} provides a possibility to choose one 
of the following initial distributions:
\begin{itemize}
 \item Power law for the surface density: $\Sigma \sim \left( \frac{h - 
h_\mathrm{in}}{h_\mathrm{out} - h_\mathrm{in}} \right)^{k_\Sigma}$.
 \item Power law for the viscous torque: $F \sim \left( \frac{h - 
h_\mathrm{in}}{h_\mathrm{out} - h_\mathrm{in}} \right)^{k_F}$.
 \item Sinus law for the viscous torque: $F \sim \sin\left( \frac{h - 
h_\mathrm{in}}{h_\mathrm{out} - h_\mathrm{in}} \frac{\piup}{2} \right)$. This 
law satisfies both border conditions for $F(h)$.
 \item {Two-parametric Gaussian distribution for the viscous torque.
}
 \item Quasi-stationary distribution: 
 $F \sim f_F\left( \frac{h}{h_\mathrm{out}} \right) \frac{1 - 
h_\mathrm{in}/h}{1 - h_\mathrm{in}/h_\mathrm{out}}$, 
 where  $f_F\left( \frac{h}{h_\mathrm{out}} \right)$ is a coordinate part of the 
self-similar analytic solution of the diffusion equation in the assumption of 
the fixed outer radius and zero inner radius \citep{lipunova_shakura2000}.
\end{itemize}
All results presented in \S~\ref{s.irradiation_dominated_evolution} are 
obtained using the quasi-stationary distribution as the initial one.

Fig.~\ref{fig.comp_opac} shows a comparison of the diffusion equation solutions 
obtained with {\sc freddi} for different analytic opacity laws and the solution 
for tabulated opacity  values~\citep[OPAL tables;][]{iglesias-rogers1996}. We alternatively use the Kramers {opacity} 
$\varkappa = 5\times10^{24} \rho/T^{7/2}$~cm$^2$~g$^{-1}$ or 
$\varkappa = 
1.5\times10^{20} \rho/T^{5/2}$~cm$^2$~g$^{-1}$ \citet{bell-lin1994}. {The initial}
power-law  distribution of the viscous torque with {an index of} $k_F = 6$ is set. The 
third solution (the solid line) is obtained with  a code described in 
\citet{malanchev-shakura2015} {using tabulated  OPAL }  opacities {and a}
quasi-stationary initial distribution. The {values of the disk parameters 
used in all of the three calculations are the same}: $\alpha = 0.5$, $m_\x = 10$, $\ak = 0$, and 
{$\rhot=10^{11}$~cm, the latter being taken constant}. {Figure}~\ref{fig.comp_opac} demonstrates that for {this}
specific problem, both the solutions with the Kramers opacity law and the approximation by 
\citet{bell-lin1994} {provide a comparable accuracy in comparison with the solution with the table OPAL opacities.}

The high speed  of {\sc freddi} and {a} specifically developed 
interface make it very useful for fitting the Shakura-Sunyaev $\alpha$-parameter and  {the} disk radius $\rhot$.
One can set {the} distance $d$,  the irradiation factor 
$C_\irr$, and {the} inclination $i$ of the system to obtain {the} X-ray flux
and the spectral flux density at any wavelength. 
It should be noted that {\sc freddi} uses {an analytical} vertical structure and thus 
works orders of magnitude faster than any {other} code, which uses tabulated opacity values 
and {numerically solving the vertical structure equations}. {\sc freddi} has several tuning 
parameters: {a} time step, {a} number of  coordinate steps and {a} type of coordinate grid 
(linear or logarithmic in terms of $h$).
 {\sc freddi} is written on \verb!C++! and {has} 
a  user-friendly command-line interface.

\bibliographystyle{mnras}
\bibliography{lipunova}
\label{lastpage}
\end{document}